\documentclass[psfig]{mn2e}
\usepackage{graphicx}
\usepackage{caption}
\usepackage{subcaption}

\begin{document}

\title[Spectroscopy of RL-QSOs \& Lyman-${\bf \alpha}$ Nebulae]
 {Spectroscopy of 7 Radio-Loud QSOs at ${\bf 2<z<6}$: Giant
Lyman-${\bf \alpha}$ Emission Nebulae Accreting onto Host Galaxies}
\author[N.Roche, A.Humphrey, L.Binette]
 {Nathan Roche$^1$, Andrew Humphrey$^{1,2}$ and Luc Binette$^{3}$\\
\thanks{nathan.roche@astro.up.pt; andrew.humphrey@astro.up.pt; lbinette@astro.unam.mx}\\
$^1$ Instituto de Astrof\'isica e Ci\^encias do Espa\c co, 
Universidade do Porto, CAUP, Rua das Estrelas, 4150-762 Porto, Portugal.\\
$^2$ Instituto Nacional de Astrof\'{i}sica, \'Optica y Electr\'onica (INAOE), Aptdo. Postal 51 y 216, 72000 Puebla, Pue., M\'exico.\\
$^3$ Instituto de Astronom\'ia, Universidad Nacional Aut\'onoma de M\'exico, Ap. 70-264, 04510 
M\'exico D.F., M\'exico.}

\bibliographystyle{unsrt} \bibliographystyle{unsrt} 

\date{11 July 2014}

\pagerange{\pageref{firstpage}--\pageref{lastpage}} \pubyear{2014}

\maketitle
 
\label{firstpage}

\begin{abstract}

We performed long-slit optical spectroscopy (Gran Telescopio Canarias - Optical System for Imaging and low Resolution Integrated Spectroscopy) of 6 radio-loud QSOs at redshifts $2<z<3$, known to have giant ($\sim 50$--100 kpc) Lyman-$\alpha$ emitting nebulae, and detect extended Lyman-$\alpha$ emission for 4,  with  surface brightness  
$\sim10^{-16}$  ergs $\rm cm^{-2}s^{-1}arcsec^{-2}$ and line full width at half-maximum 400--1100 (mean 863) km $\rm s^{-1}$.  We also observed the $z\simeq 5.9$  radio-loud QSO, SDSS J2228+0110, and find evidence of  a  $\geq 10$ kpc extended Lyman-$\alpha$ emission nebula, a new discovery for this high-redshift object. 

Spatially-resolved kinematics of the 5 nebulae are examined by fitting the  Lyman-$\alpha$  wavelength at a series of positions along the slit. We found the line-of-sight velocity $\Delta(v)$ profiles to be relatively flat. However, 3 of the nebulae appear systematically redshifted by 250--460 km $\rm s^{-1}$ relative to the Lyman-$\alpha$ line of the QSO (with no offset for the other two), which we argue is evidence for infall. One of these (Q0805+046)  had a small ($\sim 100$ km $\rm s^{-1}$) velocity shift across its diameter and a steep gradient at the centre.   Differences in line-of-sight kinematics between these 5 giant nebulae and similar nebulae associated with  high-redshift radio galaxies (which can show steep velocity gradients) may be due to  an orientation effect, which brings infall/outflow rather than rotation into greater prominence for the sources observed `on-axis' as QSOs. 
  
    \end{abstract}


\section{Introduction}

High redshift ($z>2$) QSOs (quasi-stellar objects)  and other Active Galactic Nuclei (AGN) are of 
great importance in cosmology and the study of galaxy evolution, being the most luminous of all 
long-duration sources and hosted by the most massive  galaxies existing at these redshifts. At the same redshifts, surrounding these powerful AGN or other very active galaxies, are found the uncommon and visually spectacular  `Lyman-$\alpha$ blobs' or giant high-redshift Lyman-$\alpha$ Nebulae (HzLAN).
HzLAN typically extend over diameters $\sim 50$ to at least 120 kpc and emit most strongly in the Lyman-$\alpha$ line ($1215.67\rm \AA$), with luminosities 
$\sim 10^{44}$ ergs $\rm s^{-1}$, of the same order as the line emission from QSOs or massive starburst  galaxies. As a class of sources they are diverse and may be powered by photoionization from intense bursts of star-formation, e.g. in galaxy mergers (Yajima, Li and Zhu 2013), AGN, or mixtures of both (e.g. Colbert et al. 2011).

Many of the largest and brightest HzLANs are associated with powerful high-redshift radio galaxies (HzRGs) (e.g. Ojik et al. 1997; Villar-Mart\'in et al. 2003; S\'anchez and Humphrey 2009) or radio-loud QSOs (e.g. Heckman et al. 1991a,b). In these cases, some Lyman-$\alpha$ emission could have been shock-excited by the radio jets interacting with dense clouds (also triggering star-formation) (Bicknell et al. 2000).  Indeed, there is much evidence to suggest that radio-mode feedback can have an important impact on the gas dynamics in HZRGs (e.g., Humphrey et al. 2006; Nesvadba et al. 2008). Detections of Lyman-$\alpha$ polarization in two HzLAN (Hayes et al. 2011; Humphrey et al. 2013b) imply that some fraction of the Lyman-$\alpha$ photons may be nuclear emission resonantly scattered by extended neutral hydrogen (HI), in addition to in-situ emission from the photoionized nebula.  

Some radio-quiet AGN, both QSOs and type-2, at these redshifts also have (usually slightly less luminous) HzLANs (Christensen et al. 2006; Smith et al. 2009). Most known HzLAN are at $2<z<4$, however the highest redshift examples are `Himiko' at $z=6.6$, now identified as a triple-galaxy merger undergoing starburst with $\rm star-formation rate (SFR)\sim 100~M_{\odot} yr^{-1}$ (Ouchi et al. 2009, 2013), and the $z=6.4$ quasar CFHQSJ2329-301 (Willott et al. 2011).

Radio-loud (RL-) QSOs and HzRGs, according to the unified scheme of active galaxies, are essentially the same class of  AGN viewed at different orientations: for the QSOs, along/near the radio axis, so the AGN is directly visible as an intense point-source; for the RGs, at a large or perpendicular angle, so the nucleus is obscured by a central torus but the radio jets are extended to either side. The orientation should have little or no effect  on the visibility of a giant Lyman-$\alpha$ nebula (in e.g. narrow-band imaging), as this is far larger than the AGN and torus structure.

Heckman et al. (1991a, hereafter H91) obtained Lyman-$\alpha$ and broad-band imaging on a sample of 18 RL-QSOs at $z>2$ and found 15 to have extended Lyman-$\alpha$ emission from nebulae, with diameters $\sim 11$ arcsec $\simeq 90$ kpc, surface brightness $\sim 10^{-16}$ ergs $\rm cm^{-2} s^{-1} arcsec^{-2}$, fluxes $F_{Ly\alpha}\sim 10^{-15}$ to $10^{-14}$ ergs $\rm cm^{-2} s^{-1}$, and hence $L_{Ly\alpha}\sim 10^{44}$ ergs $\rm s^{-1}$,  clearly putting them in the class of HzLAN. As these galaxies are QSOs, their total UV flux is  dominated by the central point-source, but the HzLANs were bright enough to contribute $\sim 10$--$30\%$ in Lyman-$\alpha$. Although some HzLAN were too round to define a position angle, for most the long axis was aligned within $<30$ deg of the radio axis, with the brightest Lyman-$\alpha$ and radio emission coinciding (on the same side of the AGN). Heckman et al. (1991b) took long-slit optical spectroscopy 
(with the Kitt Peak National Observatory 4m Mayall telescope) for 5 of the H91 RL-QSOs, and found the nebular Lyman-$\alpha$ lines to have large velocity dispersions of $\rm FWHM\sim 1000$--1500 km $\rm s^{-1}$; they did not detect velocity gradients across the nebulae, but their relatively low spectral resolution of $R\simeq 450$ would not have well-constrained this. 

Van Ojik et al. (1997) obtained higher-resolution long-slit spectra of 17 HzRGs and one RL-QSO  with extended Lyman-$\alpha$ emission (2--17 arcsec) at $2<z<3$. The majority (11) were found to show one or more narrow absorption lines within the broader ($\rm FWHM\simeq 700$--1600 km $\rm s^{-1}$) Lyman-$\alpha$ emission. These narrow, strong absorption lines in most cases covered the whole spatial extent of the nebula, with only small velocity offsets from the QSO emission (mostly $<250$ km $\rm s^{-1}$), implying  the absorbing structures were large and physically associated with the radio galaxies. Binette et al. (2000) proposed that absorption seen in both Lyman-$\alpha$ and CIV in a HzRG spectrum was caused by a large shell of low-metallicity and low-density gas, even further out from the galaxy than the emitting nebula and the radio jets, perhaps a vestige from galaxy formation and predating the AGN activity. Humphrey et al. (2008), with integral-field and long-slit spectroscopy of another HzRG at $z=2.63$, found a double set of absorption lines at large blueshifts  $\simeq 600$ and 2000 km $\rm s^{-1}$, in Lyman-$\alpha$ and several other lines such as SiIV(1393,1402), again covering the whole emission nebula.

Villar-Mart\'in et al. (2003), using long-slit spectroscopy (Keck-Low Resolution Imaging Spectrometer, $R\simeq 2000$) of 10 radio galaxies at $2.2<z<3$, found the spectra of the giant nebulae to show,  in addition to the strong Lyman-$\alpha$, HeII1640 and in some cases CIV1550 and/or NV1240 emission lines. These nebula lines showed  relatively moderate velocity dispersions ($\rm FWHM\simeq 250$--800 km $\rm s^{-1}$), and in most cases significant velocity shifts of a few hundred km $\rm s^{-1}$ across the $\sim 100$ kpc nebula diameter. The nebulae were described as `halos of kinematically quiescent gas', and as `a common ingredient of high redshift radio galaxies'.

Integral-field spectroscopy (Very Large Telescope - Visible Multi Object Spectrograph, VLT-VIMOS) of three $z\sim 2.5$ radio galaxies (Villar-Mart\'in et al. 2007) found the nebular Lyman-$\alpha$ emission to be aligned with the radio axes, and in two examples double-peaked, with radial-velocity offsets between the components of a few hundred km $\rm s^{-1}$ (suggestive of rotating systems), while in the third HzLAN the centre was redshifted relative to the edges (a pattern suggestive of infall).  Humphrey at al. (2007) described an interesting correlation seen in almost all radio galaxies with HzLAN -- the more redshifted side of the nebula is the brighter in both Lyman-$\alpha$ and radio flux. The interpretation was that this was the near-side relative to the observer -- so here the radio jet is approaching and is Doppler-boosted while the Lyman-$\alpha$ is less obscured. The relative redshift of the near side would then imply the  emitting nebula has  a significant component of infall towards the nucleus (whether or not it  is also rotating about it).
 Adams,  Hill and MacQueen (2009) fit their observations of a bimodal Lyman-$\alpha$ profile in one $z=3.4$ RG with a somewhat similar model of an photoionized bicone within a massive, infalling and resonantly scattering HI halo.  S\'anchez and Humphrey (2009), with integral-field spectroscopy, found two more HzRG giant nebulae to emit in CIV and HeII1640 as well as Lyman-$\alpha$ (confirming the gas is photoionized by the obscured AGN); one (4C+40.36) showed a rather extreme velocity gradient, $\Delta(v)\simeq 1000$ km $\rm  s^{-1}$ across the nebula, which is more than expected from rotation alone and most easily explained by infall/outflow.

Humphrey et al (2013) observed the one RL-QSO/HzLAN in the van Ojik et al. (1997) sample, 
TXS1436+157 at $z=2.54$, with the new OSIRIS long-slit spectrograph on the 10.4m Gran Telescopio Canarias (GTC). The choice of QSO rather than radio galaxy was motivated by the direct view of the intense AGN light against which absorption would more sensitively be revealed, and also for a comparison of RL-QSO nebulae with HzRGs -- if the nebulae are aligned  with the radio axes, the different orientation of these two classes of source is likely to influence how the internal velocity structure of the nebula appears when viewed along the line-of sight as an emission-line velocity map.  For TXS1436+157 a strong absorption line is seen within the broader Lyman-$\alpha$ emission, mildly blueshifted ($\sim 100$ km $\rm s^{-1}$) and partially absorbing over the whole nebula. This was interpreted as a large (radius $\geq 40$ kpc) and massive ($>10^{11}\rm M_{\odot}$) shell of hydrogen (with small amounts of heavy elements), formed where an expanding super bubble of hot gas, driven outward by supernovae following a starburst in the central galaxy, encounters and sweeps up a low-metallicity intergalactic medium (expanding superbubbles around radio galaxies have also been seen in [OIII]5007 emission; Humphrey et al. 2009).  In contrast, the HzLAN in TXS1436+157 is slightly redshifted (110 km $\rm s^{-1}$) relative to the AGN, suggesting infall.

Further GTC-OSIRIS long-slit spectroscopy was proposed, targeting a larger sample of  $2<z<3$ RL-QSOs from H91, to study at $R\geq 2000$ spectral resolution both the extended Lyman-$\alpha$ emission and intrinsic absorbers.  So far we have observed 6 of these galaxies, and present the first results in this paper. Additionally we observed at similar resolution a much fainter RL-QSO, one of the most distant known, discovered by Zeimann et al. (2011) at $z=5.95$. 

Section 2 of this paper lists the observed galaxies, and describes the observations and data reduction.  Section 3 presents our spectra for the 7 RL-QSOs and extended nebulae.   Section 4 is a further analysis of the spatially-resolved kinematics of the nebulae, using the Lyman-$\alpha$ profiles. Our findings are discussed and compared with previous observations of radio galaxies and QSOs in Section 5, and finally Section 6 summarizes the main conclusions. 
 
 Magnitudes are given in the AB system where $m_{AB}=-48.60-2.5$ log $F_{\nu}$ (in ergs $\rm cm^{-2}s^{-1}Hz^{-1}$). We assume throughout a spatially flat cosmology with $H_0=70$ km $\rm s^{-1}Mpc^{-1}$, $\Omega_{M}=0.27$ and $\Omega_{\Lambda}=0.73$, for which the age of the 
Universe is 13.86 Gyr and 1 arcsec is (8.596,7.942,5.916) kpc at z=(2,3,6). 
\section{Data}
\subsection{Sample}
Our spectroscopy sample consists of 6 of the H91 $2<z<3$ RL-QSOs already known to have giant nebulae, plus the $z\sim 6$ QSO SDSS J2228$+$0110 (Zeimann et al. 2011), discovered in the Sloan Digital Sky Survey (SDSS) deep Stripe 82 (matched with a VLA radio survey). It is much fainter than the first six ($m_z=22.3$ compared to $m_B\simeq 19$), and like almost all QSOs has a strong Lyman-$\alpha$ emission line, but little else is known. Table 1 lists the 7 targets, with positions and redshifts.

The H91 RL-QSOs were originally selected from a larger sample of 105 $z>1.5$ QSOs observed by Barthel et al. (1988) in the radio (6cm/4.9GHz) with the Very Large Array (VLA), giving 0.5-arcsec resolution radio maps. The distant SDSSJ2228+0110 has only a single, much fainter, flux measurement of 0.31mJy at 1.4 GHz (Zeimann et al. 2011), but this is still sufficient to be defined as a RL-QSO with a radio loudness index $R\simeq 1100$. 

The Heckman et al. (1991b) optical spectroscopy of 5 of the H91 QSOs included two of our sample,  Q0758+120 and Q0805+046. Using parallel apertures, they extract separate spectra  (with a relatively low resolution $R\simeq 450$) for both AGN and the associated nebulae. We shall compare our results, and with the higher-resolution OSIRIS observations will be able more sensitively characterise the kinematics. Also  Lehnert and Becker (1998) took long-slit Keck spectroscopy of Q2338+042 and its nebula.  Lehnert et al. (1999) imaged 5 RL-QSOs, including Q1658+575 and Q2338+042, with HST WFPC2 in the broad-band F555W and in narrow rest frame Lyman-$\alpha$ bands, which showed the brighter regions of the nebulae at much higher resolution.

\subsection{GTC-OSIRIS Observations}
The observations presented here were all obtained using OSIRIS (Optical System for Imaging and low Resolution Integrated Spectroscopy) on the 10.4m GTC as part of observing proposals GTC3-11AMEX (May/June/July 2011) and GTC2-11BMEX (October 2011). Table 1 details the observations, which were taken in service mode. The instrument was operated in long-slit spectroscopy mode with 100 kHz CCD readout (see OSIRIS manual). The VPH(volume-phased holographic) gratings were chosen to best observe Lyman-$\alpha$ at the galaxy redshift with sufficient spectral resolution (the grating numbers refer to lines/mm): R2500U covers 3440--4610$\rm \AA$ with resolution $R=2555$, R2000B covers 3950--$5700\rm \AA$  with $R=2165$, and 
R2500I  covers 7330--$10000 \rm \AA$ with  $R=2503$. Observations for galaxies   Q1658+575  and Q2338+042 were flagged as `attempted' by the instrument operator  indicating observing conditions were to some degree non-optimal.

\begin{table*}
\begin{tabular}{lcccccccccc}
\hline
QSO & R.A. & Dec & Redshift & Date obs. & Total $t_{exp}$ & $N_{exp}$ & Grating & PA & width & seeing \\
  &    &  & (NED)   &  (2011)  & (sec)   &  &   &  deg  & arcsec & arcsec \\
  \hline
 Q1354+258 & 209.276879  & 25.624711  &  2.0013 & 27 Jul &  2436 & 3 & R2500U & -45 & 1.0 & 1.0\\
 Q1658+575 &  254.940530  & 57.525565  & 2.173 & 5 May/2 Jun  & 4060 & 5 & R2500U & 20 & 1.0 & 1.2\\
 Q0758+120 & 120.252016  & 11.889886 &  2.6716 & 23/24 Oct & 3200 & 4 & R2000B & -55 & 0.8 & 1.2\\
Q0805+046ÊÊ&  121.9897438  & 4.5429253 &  2.877 & 25 Oct   &   2400 & 3  & R2000B &  -55  & 0.8 & 1.0\\
Q2338+042 & 355.2415218  & 4.5210618 &  2.594 & 29 Oct & 2400 & 3 & R2000B & -50 & 0.8 & 1.3\\
Q2222+051 &  336.311068   & 5.452616 & 2.323 &  3 Jun  & 2436 & 3 & R2000B & 35 & 0.8 & 1.2\\
J2228+0110 & 337.181250  &  1.175611 & 5.95 & 23 Jul & 2340 & 3 & R2500I & -52 & 0.6 & 0.8 \\
\hline
\end{tabular}
\caption{GTC-OSIRIS observations of  the 7 radio-loud QSOs: Right Ascension and Declination in degrees (J2000 system); redshifts (RA/Dec/$z$ as given on the NASA/IPAC Extragalactic Database ned.ipac.caltech.edu); date of our observations; total integrated exposure time on each source (divided into $N_{exp}$ exposures of equal length); grating; slit  position angle (PA) in degrees East of North (i.e. $\rm N=0$, $\rm E=90$, anticlockwise is positive, PA=IPA-60,54); slit width; seeing FWHM as estimated from spatial FWHM of the continuum in the reduced spectra.}
\end{table*}

\subsection{Data Reduction}
Data reduction was carried out using the standard {\sevensize IRAF} package. 
With OSIRIS operated in $2\times2$ pixel binning, each spatial pixel in our data is 0.254 arcsec, and the spectral dispersion is $\simeq$ (0.62,0.86,1.36) 
$\rm \AA$ $\rm pixel^{-1}$ for (R2500U, R2000B, and R2500I ).  Firstly, bias frames (median combined for each night) were subtracted from all images. Flat fields (using dome and lamp), were combined to a single flat for each night, normalized to a mean of unity by dividing out by a fit to the spectral-axis dependence, using {\sevensize IRAF} `flat1d').  After flat-fielding and, where necessary, spatial registration, the 3, 4 or 5 exposures for each object were combined (with {\sevensize IRAF} `crrej' cosmic ray removal), wavelength calibrated using the arc lamp spectra, and rebinned to a linear wavelength scale. The sky background and lines were subtracted (using {\sevensize IRAF }`background'). Contemporaneous exposures on spectrophotometric standard stars (Ross 640, Feige 110, G157-34, or Feige 34) were processed in the same way to obtain spectra, and, with standard {\sevensize IRAF} techniques,  define sensitivity functions for each of the observed targets. Each spectrum could then be flux-calibrated in $F_{\lambda}$, units ergs $\rm cm^{-2}s^{-1}\AA^{-1}$. We correct for Galactic reddening using {\sevensize IRAF} `deredden' and the Galactic $A_V$ values from the NED database (`New Landolt'), which in the order of Table 1 are 
$A_V=(0.047, 0.060, 0.070,0.078, 0.219,0.421,0.185)$ magnitudes.

 One-dimensional spectra  were extracted (using {\sevensize IRAF} `apall') from the 2D spectra. These spectra, in apertures fitted to the peaks in flux, will essentially be for the AGN and host galaxies. For associated extended nebulae visible on the 2D spectra, we also extract 1D spectra in parallel (close but not overlapping) apertures.
 
\section{Results}
\begin{figure}
(a)
\begin{subfigure}{0.25\textwidth}
\centering
\includegraphics[width=0.75\hsize,angle=0]{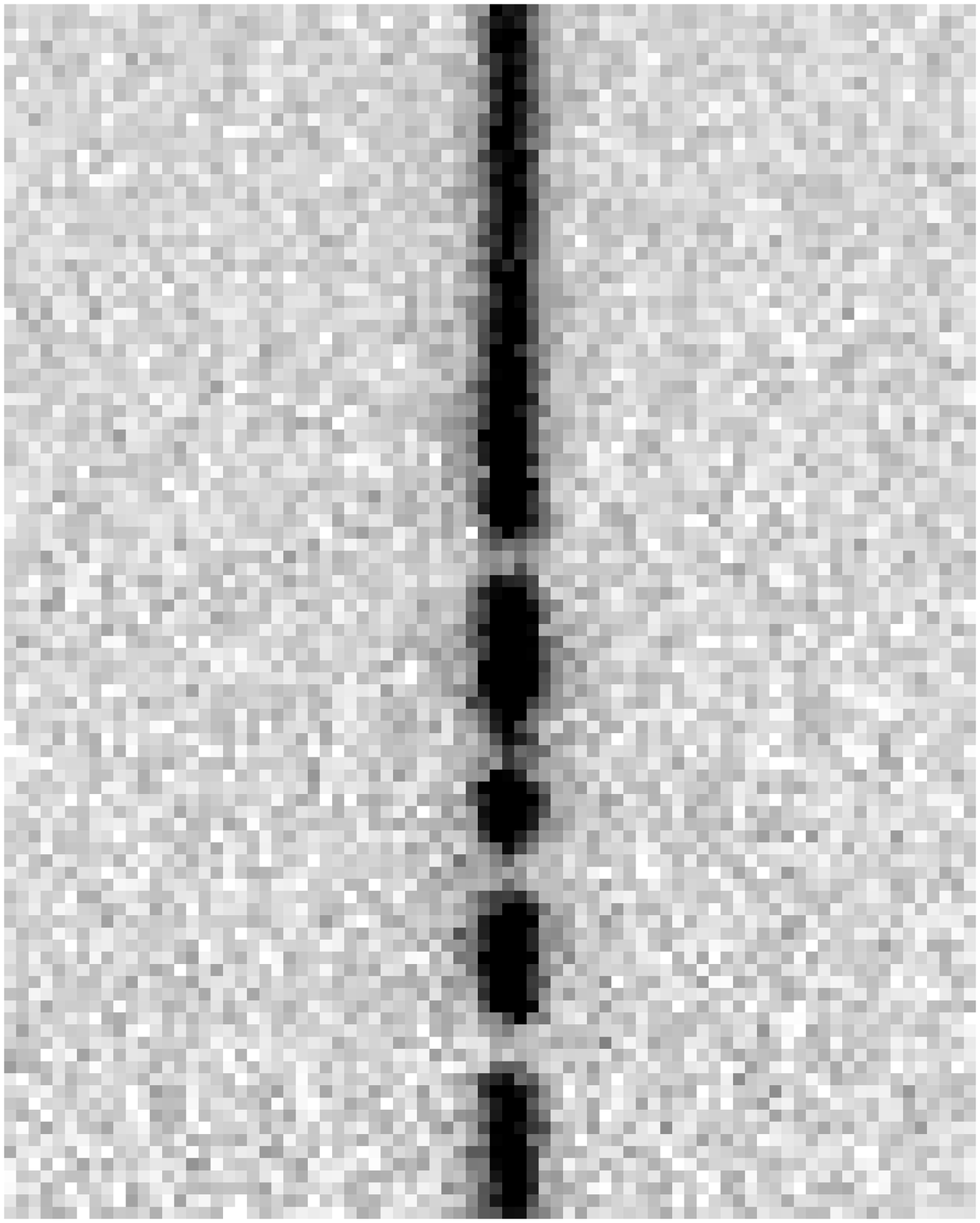}
\end{subfigure}%
\begin{subfigure}{0.25\textwidth}
\centering
\includegraphics[width=0.75\hsize,angle=0]{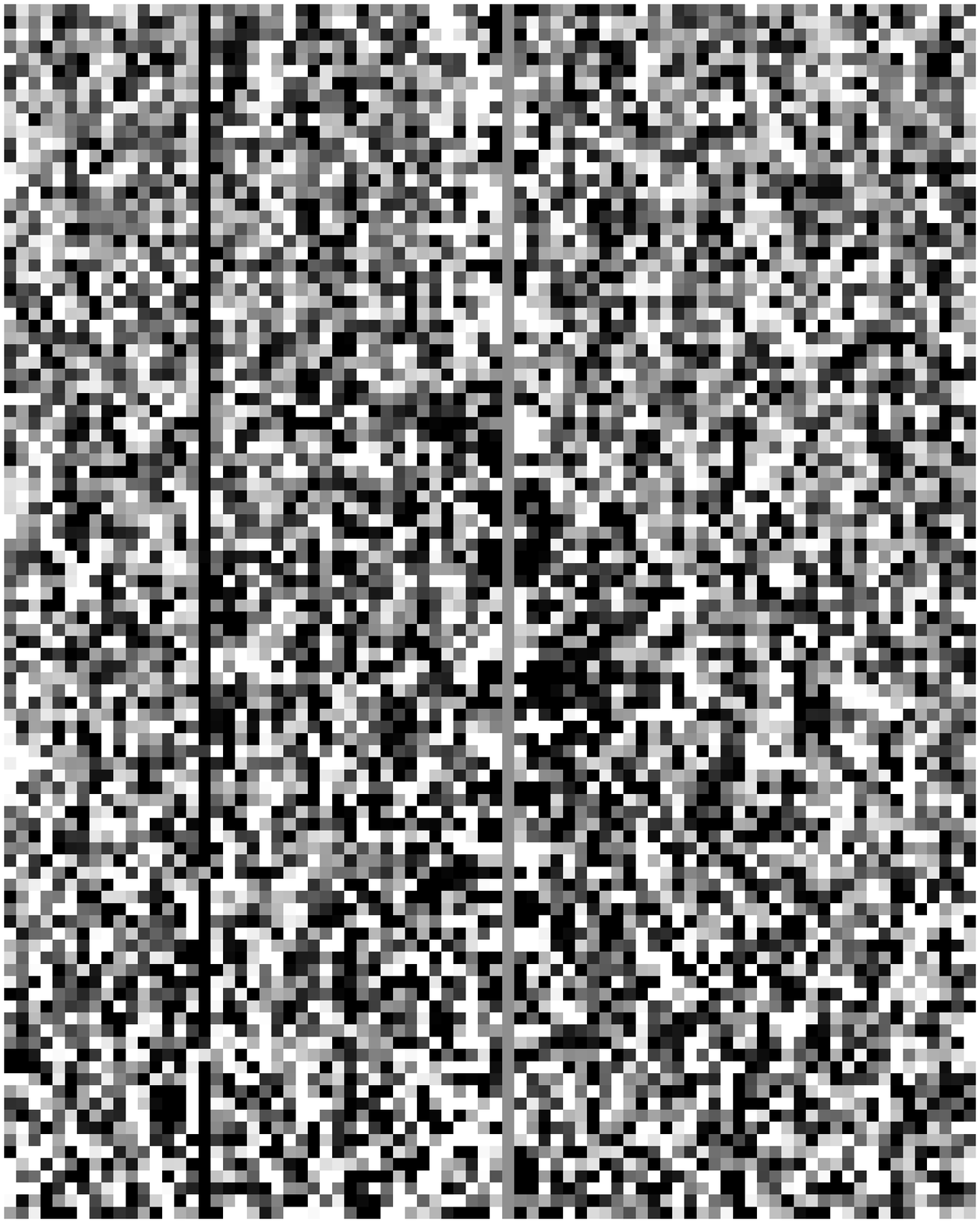}
\end{subfigure}

(b)
\begin{subfigure}{0.25\textwidth}
\centering
\includegraphics[width=0.75\hsize,angle=0]{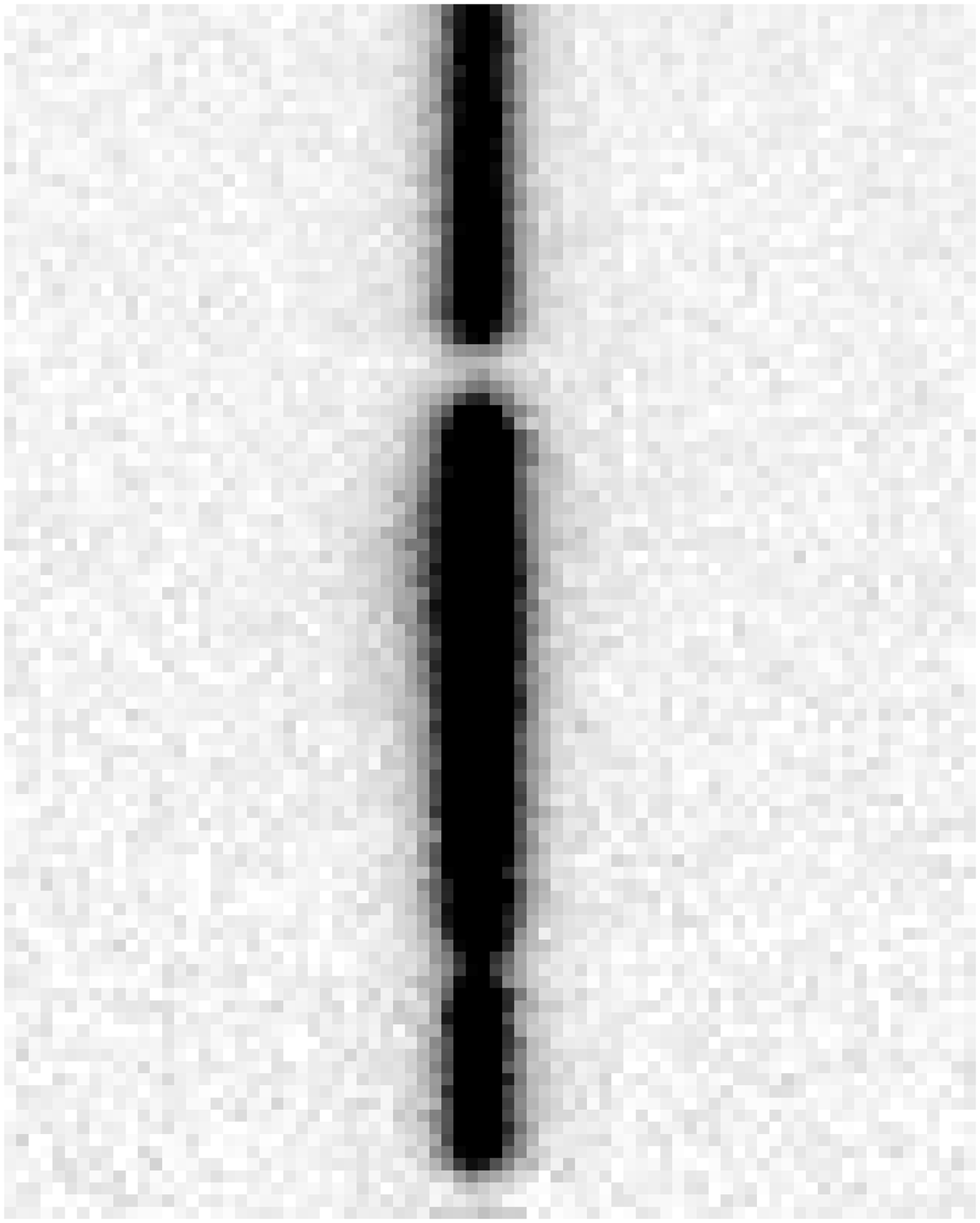}
\end{subfigure}%
\begin{subfigure}{0.25\textwidth}
\centering
\includegraphics[width=0.75\hsize,angle=0]{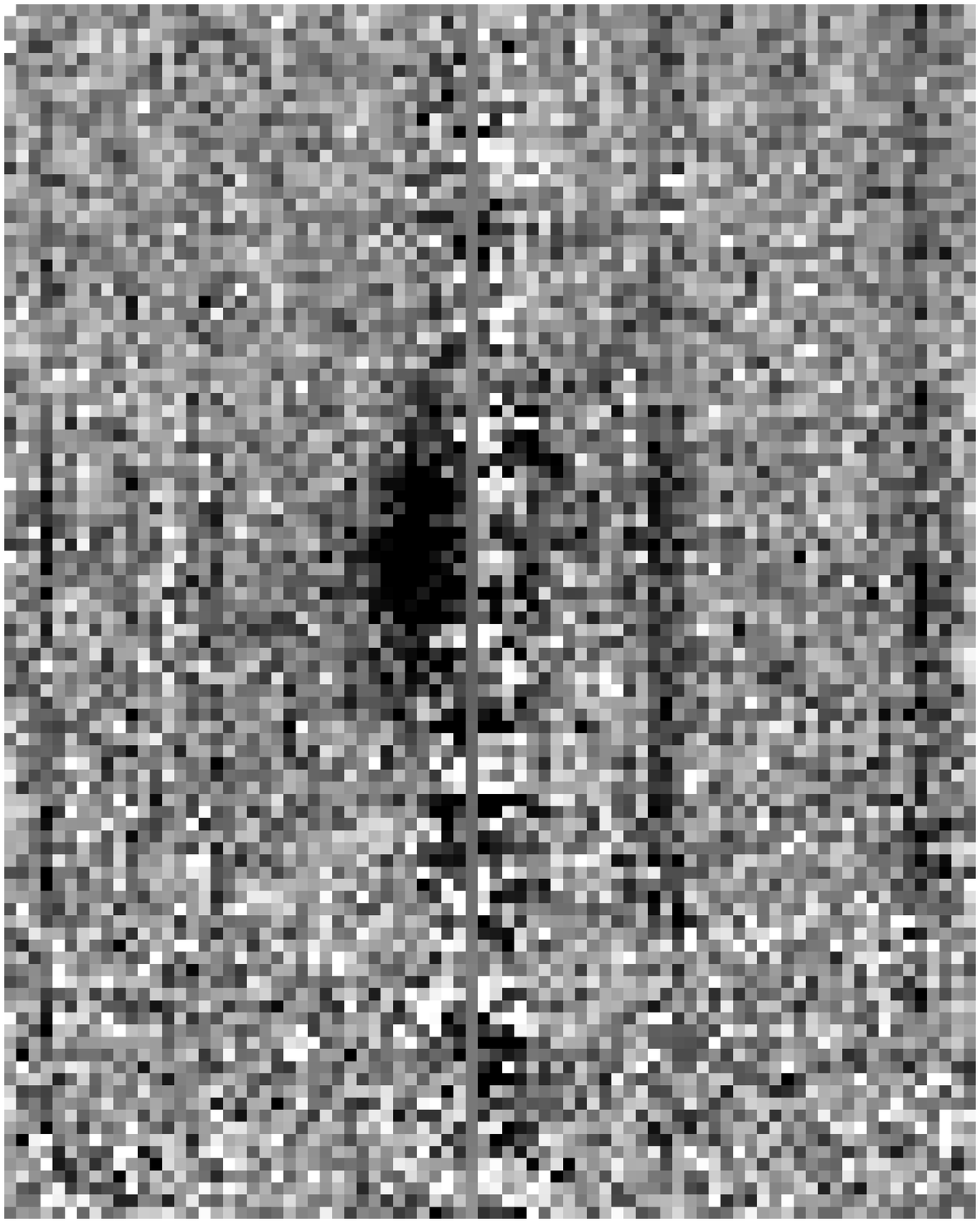}
\end{subfigure}

(c)
\begin{subfigure}{0.25\textwidth}
\centering
\includegraphics[width=0.75\hsize,angle=0]{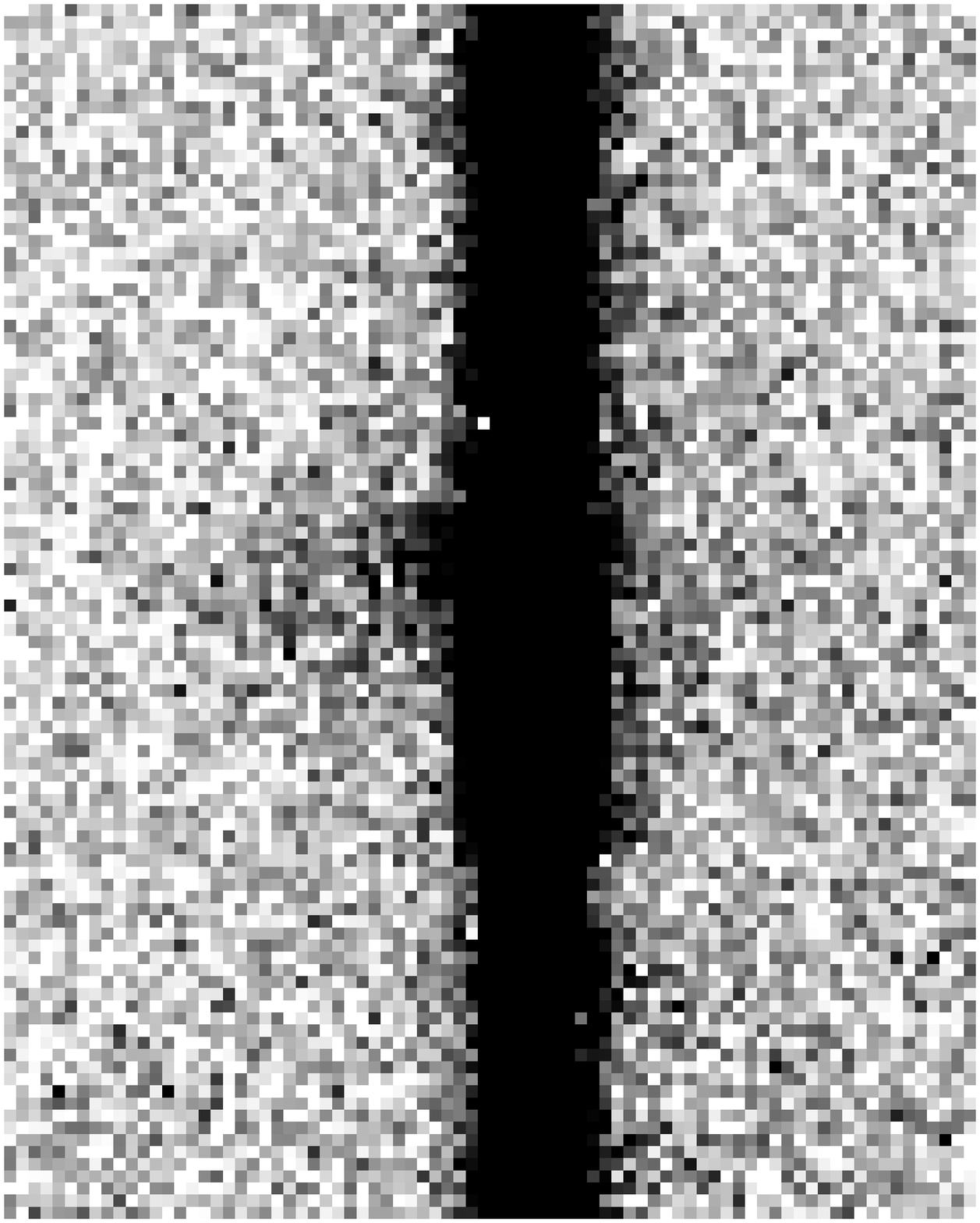}
\end{subfigure}%
\begin{subfigure}{0.25\textwidth}
\centering
\includegraphics[width=0.75\hsize,angle=0]{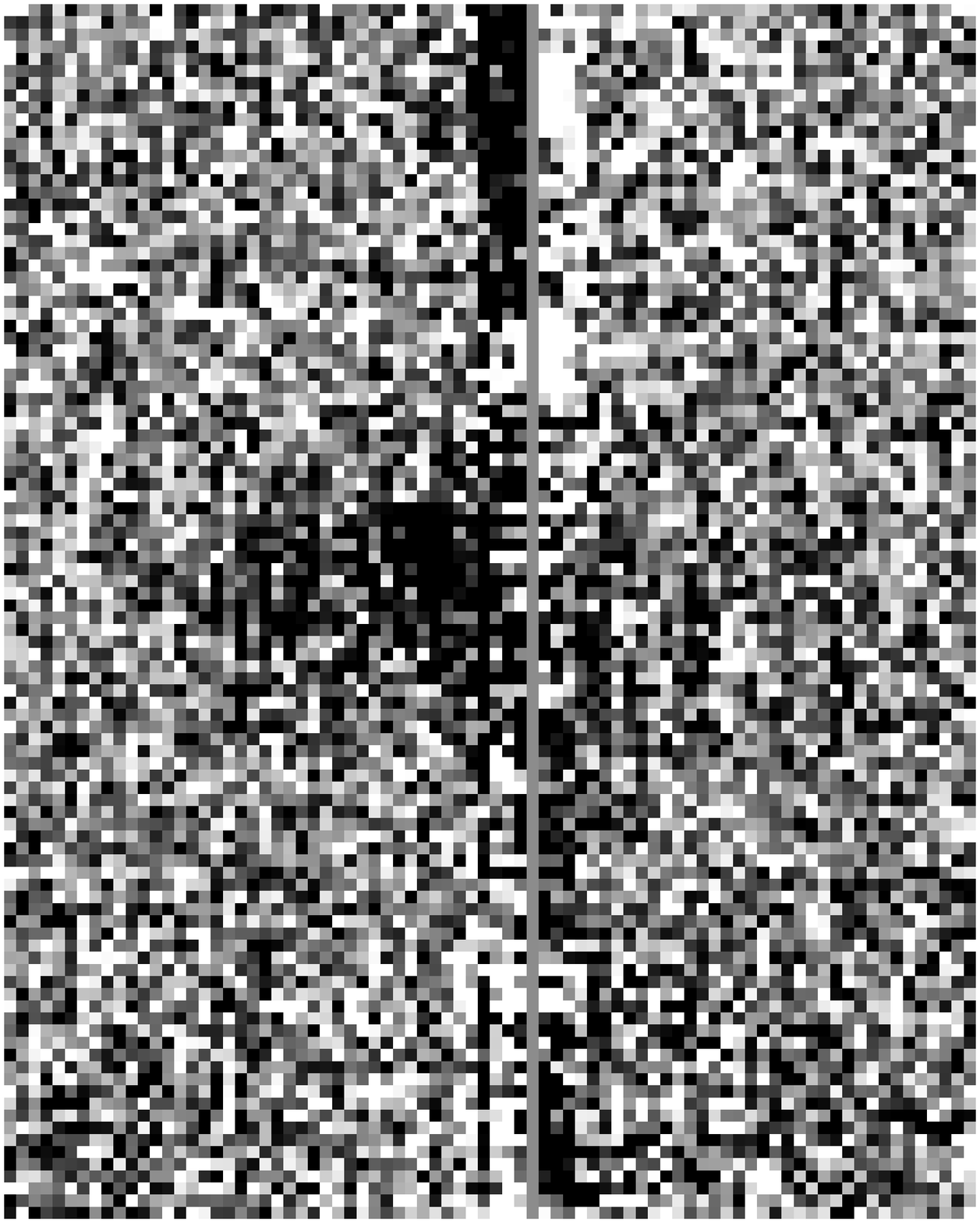}
\end{subfigure}

(d)
\begin{subfigure}{0.25\textwidth}
\centering
\includegraphics[width=0.75\hsize,angle=0]{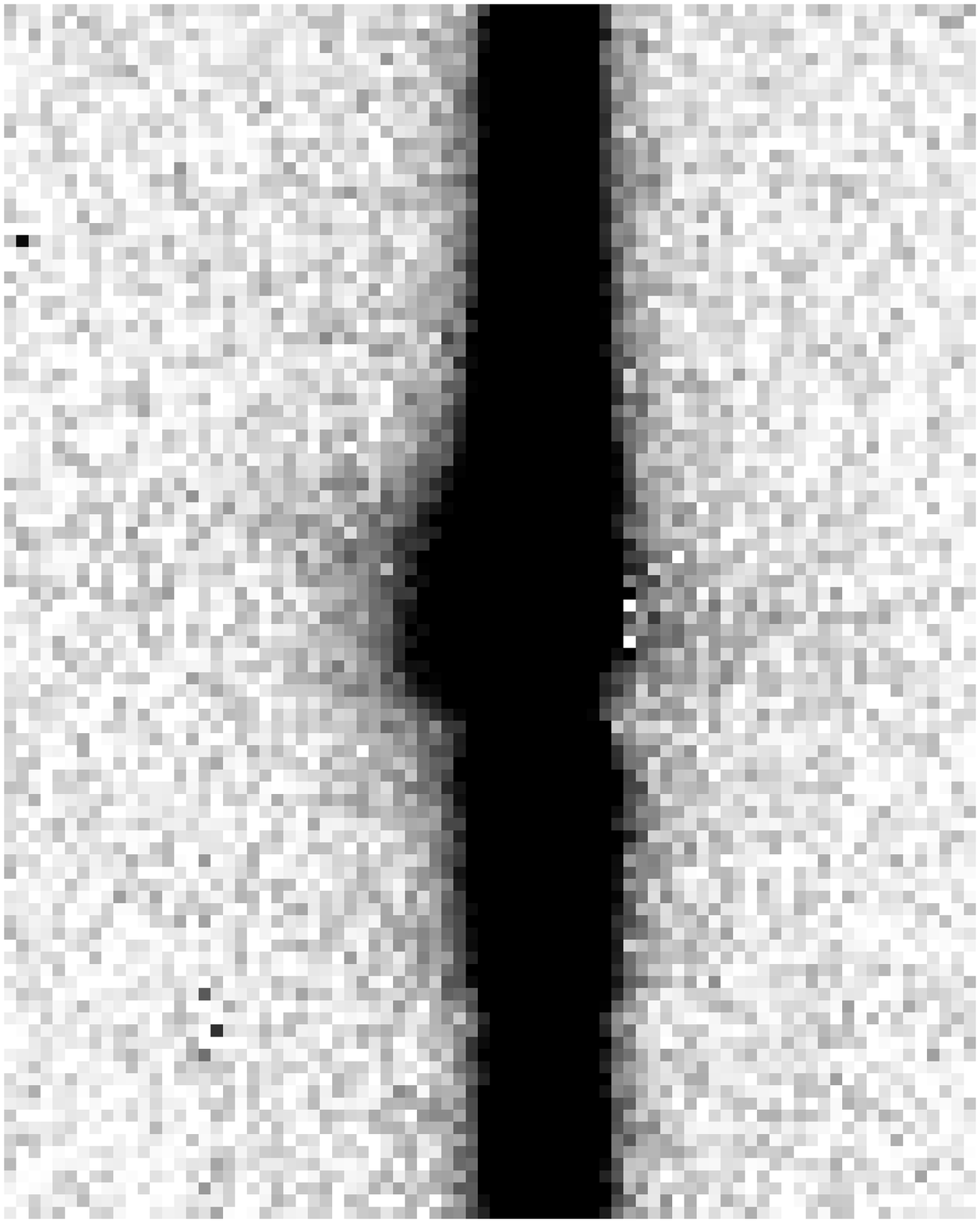}
\end{subfigure}%
\begin{subfigure}{0.25\textwidth}
\centering
\includegraphics[width=0.75\hsize,angle=0]{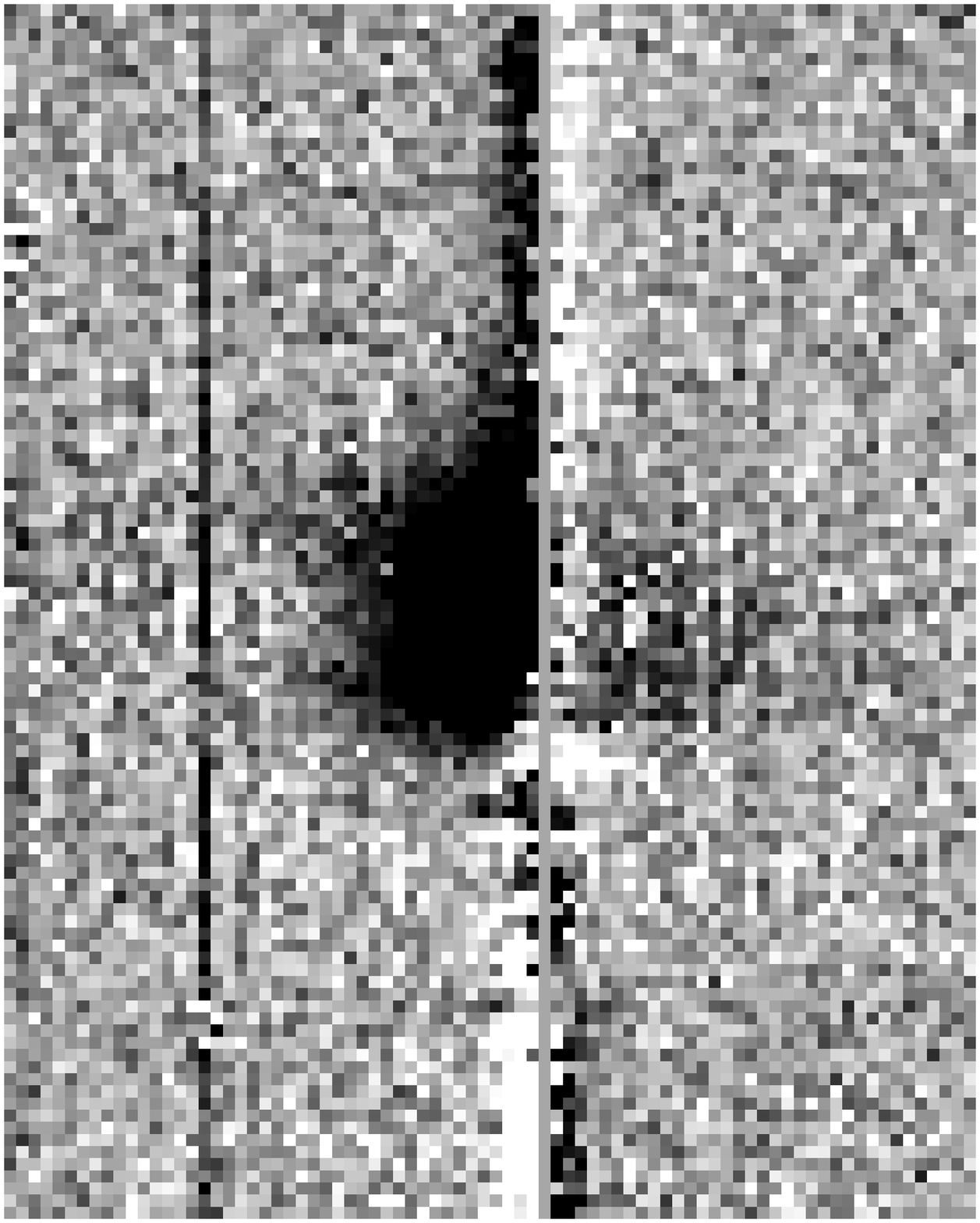}
\end{subfigure}

(e)
\begin{subfigure}{0.25\textwidth}
\centering
\includegraphics[width=0.75\hsize,angle=0]{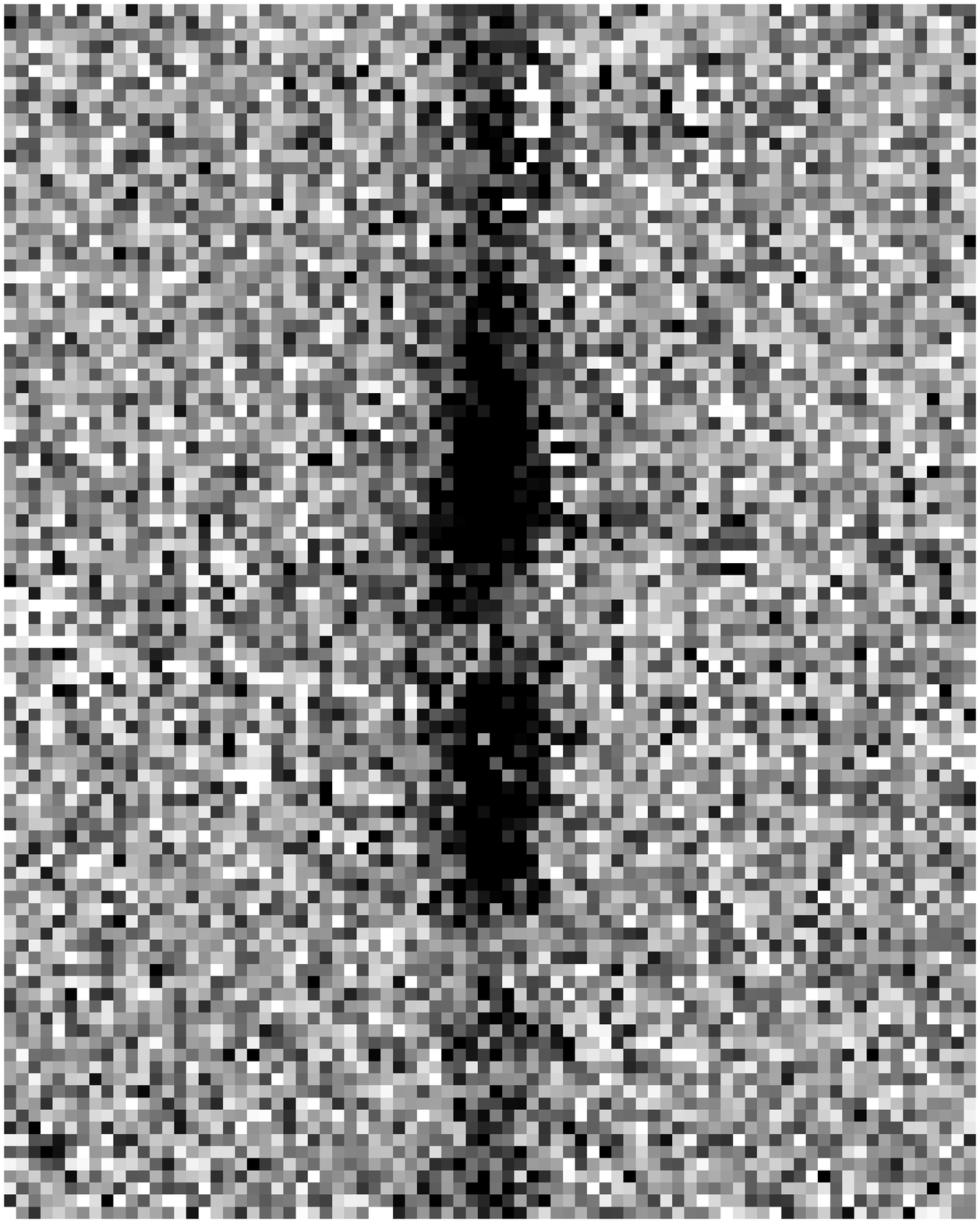}
\end{subfigure}%
\begin{subfigure}{0.25\textwidth}
\centering
\includegraphics[width=0.75\hsize,angle=0]{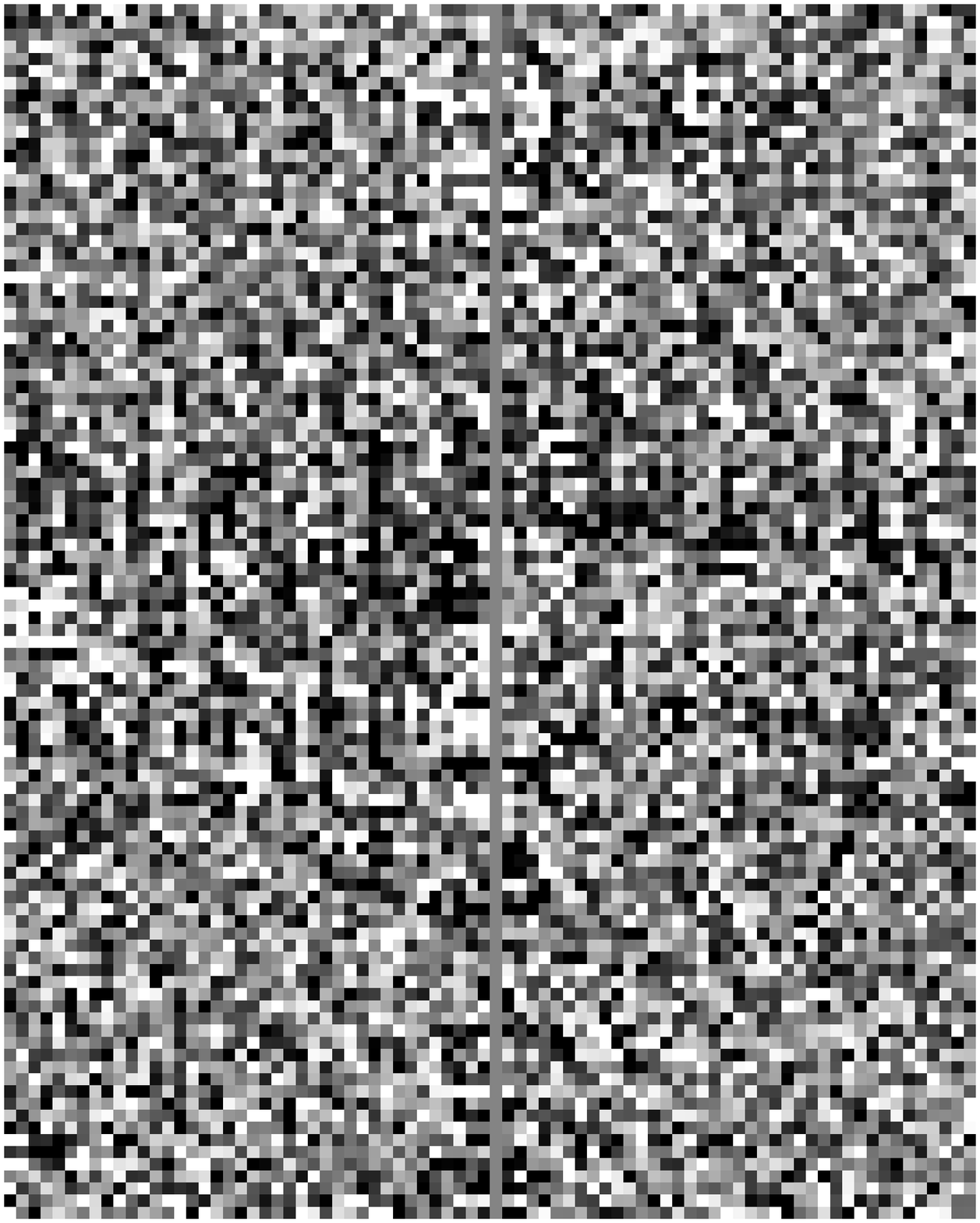}
\end{subfigure}
\end{figure}

\begin{figure}
(f)
\begin{subfigure}{0.25\textwidth}
\centering
\includegraphics[width=0.75\hsize,angle=0]{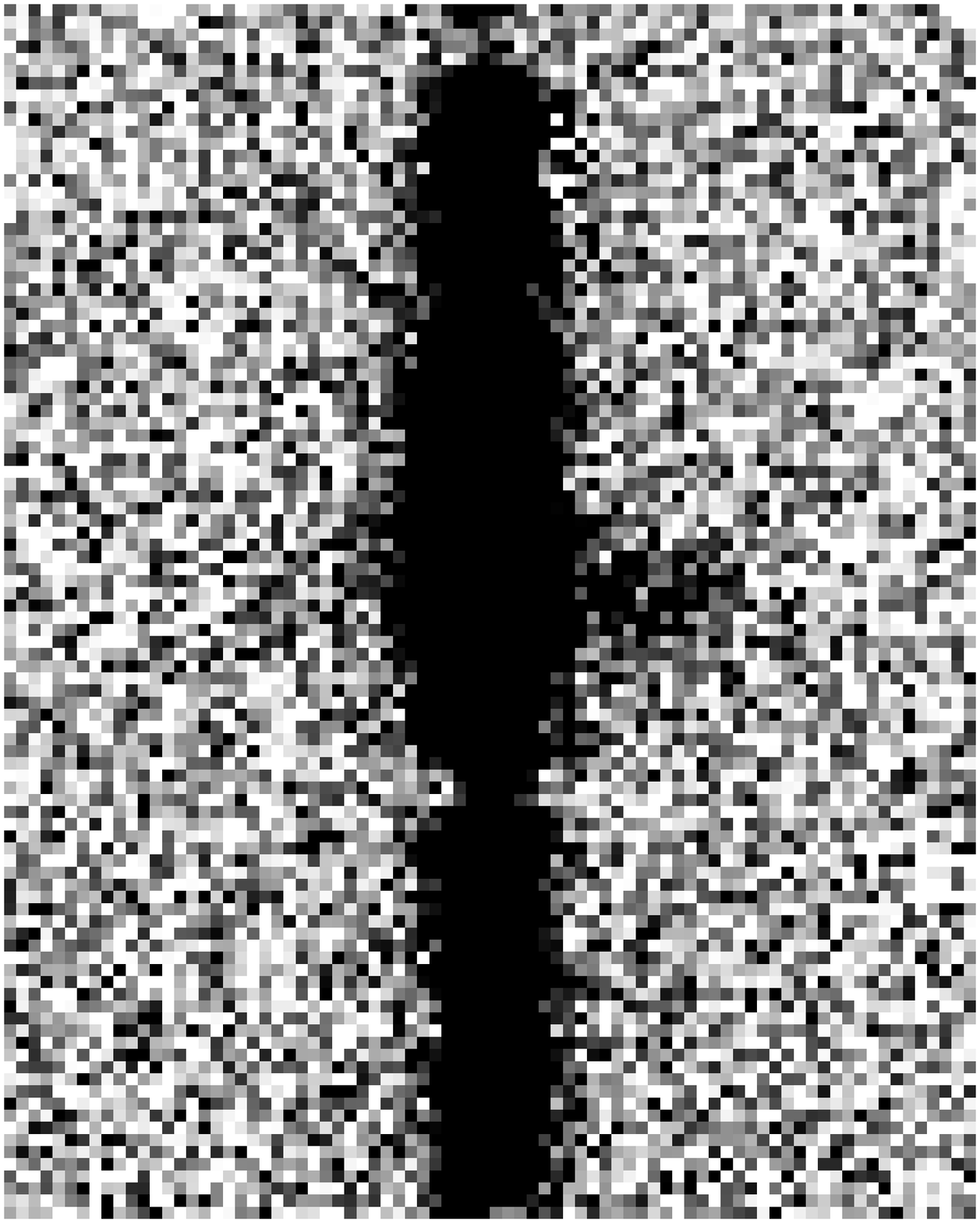}
\end{subfigure}%
\begin{subfigure}{0.25\textwidth}
\centering
\includegraphics[width=0.75\hsize,angle=0]{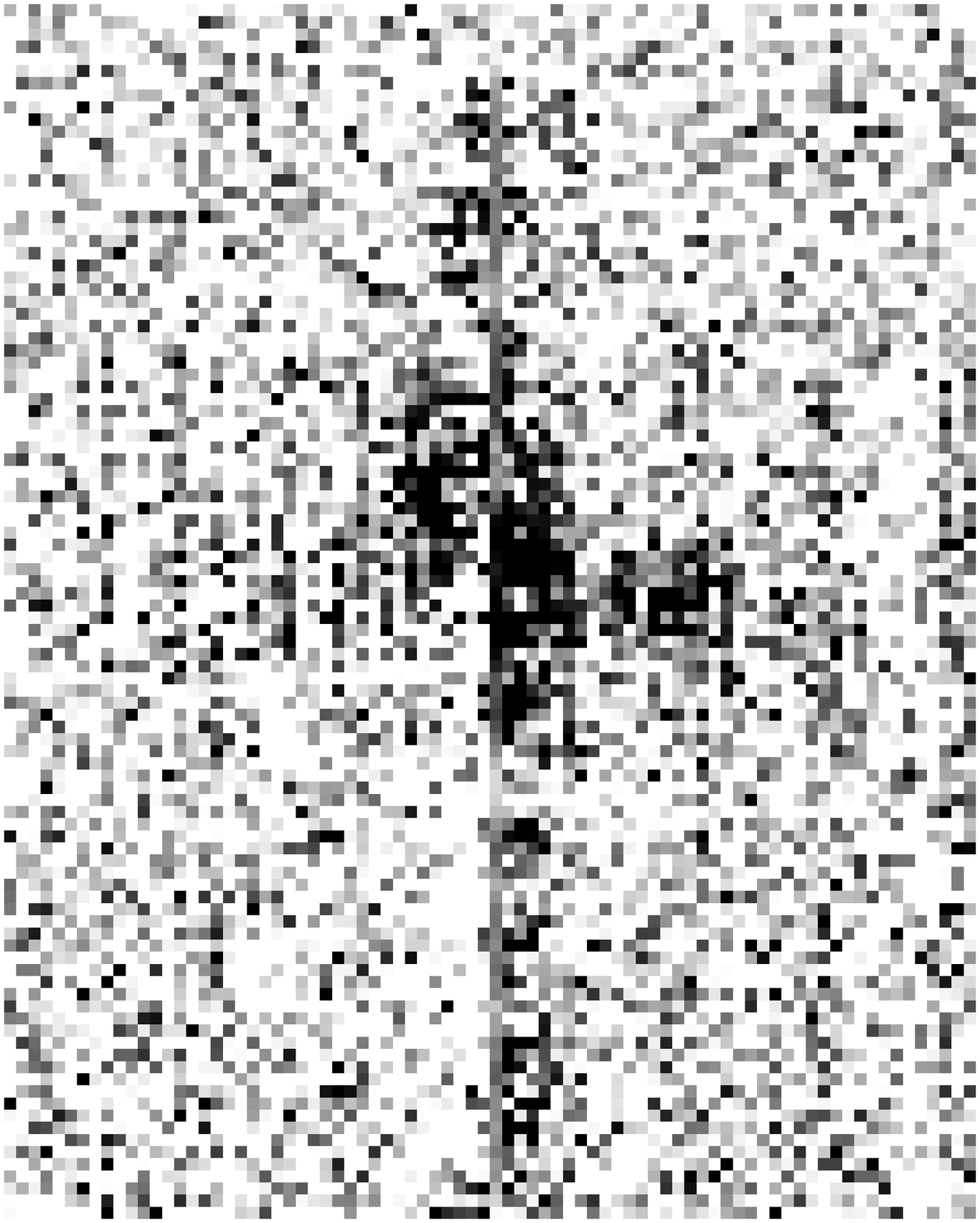}
\end{subfigure}

(g)
\begin{subfigure}{0.25\textwidth}
\centering
\includegraphics[width=0.75\hsize,angle=0]{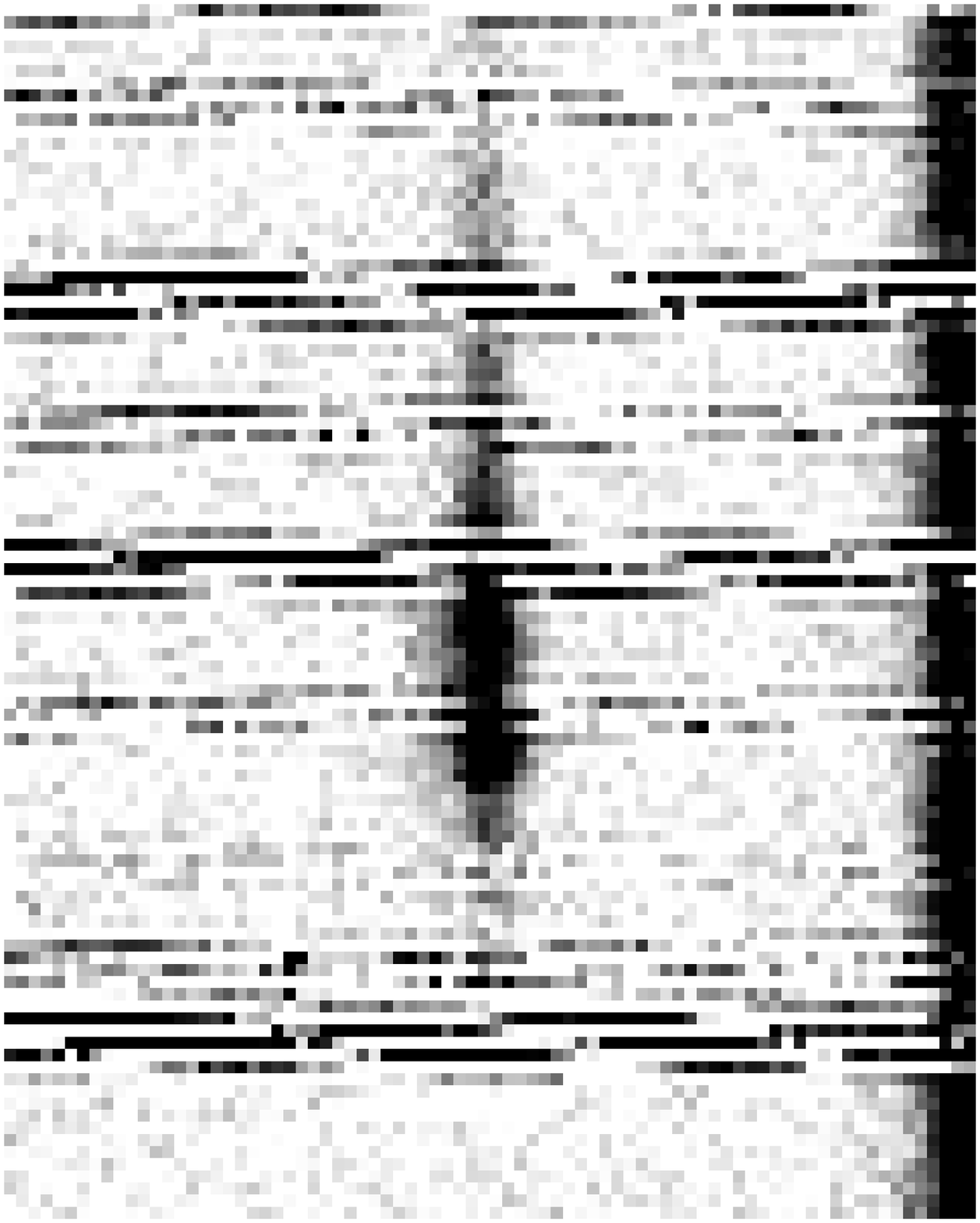}
\end{subfigure}%
\begin{subfigure}{0.25\textwidth}
\centering
\includegraphics[width=0.75\hsize,angle=0]{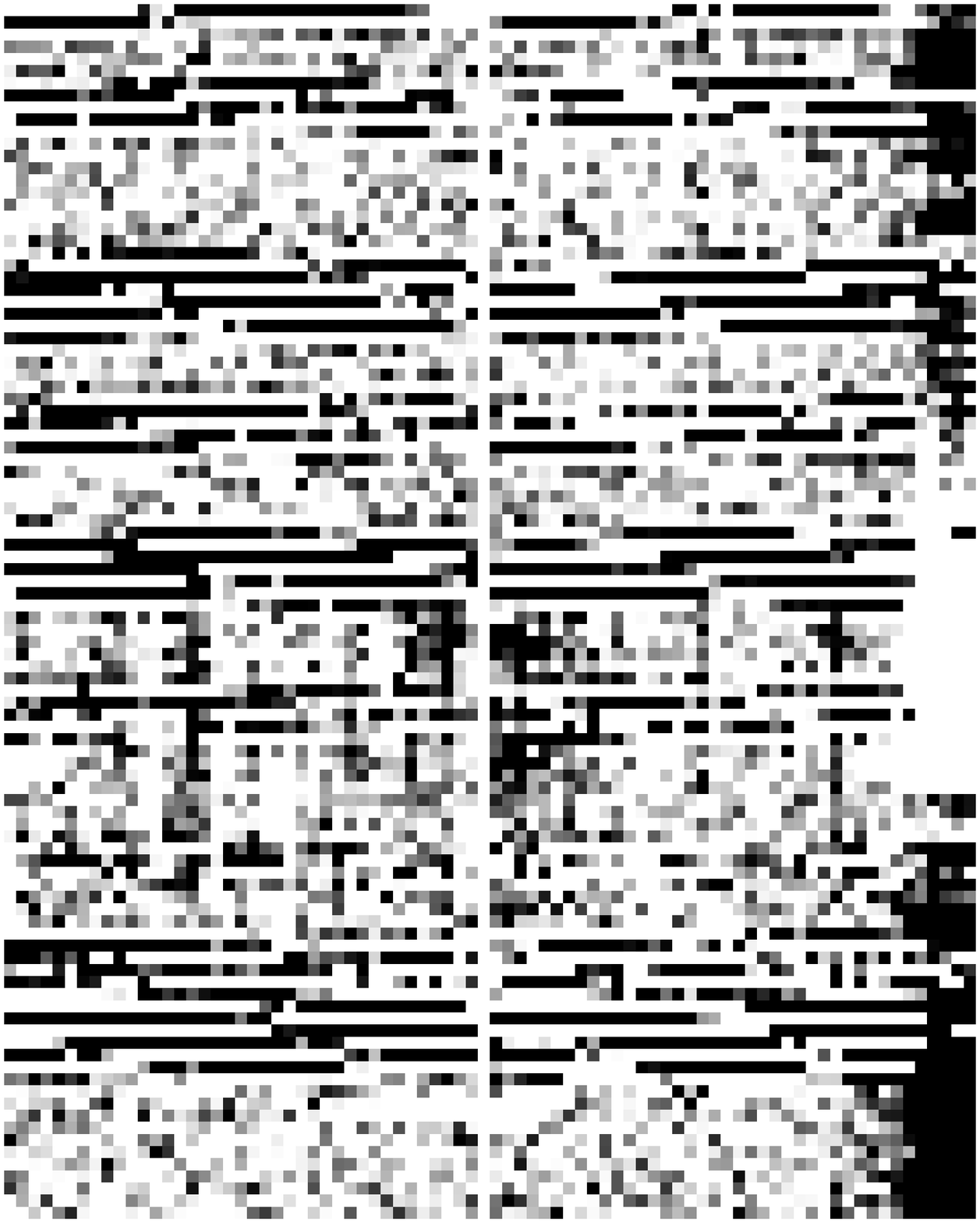}
\end{subfigure}
\caption{IRAF images of the flux-calibrated, sky-subtraced  2D spectra (width 80 pixels=20.32 arcsec) zoomed in on the   Lyman-$\alpha$ emission line (redwards is up). Each is shown as a pair of spectra; as observed (left panel) and with subtraction of the QSO/galaxy spectrum to leave only the extended nebula (right panel). Showing  (a) Q1354+258, (b) Q1658+575, (c) Q0758+120, (d) Q0805+046, (e) Q2338+042, (f) Q2222+051 and (g) SDSSJ2228+0110.} 
\end{figure}

Firstly, a strong Lyman-$\alpha$ emission line is seen on  all 7 of the quasar spectra. Fig 1 shows the Lyman-$\alpha$ regions of the 2D spectra. In addition, we measure the QSO spectrum PSF in spectral regions slightly blueward and redward of the Lyman-$\alpha$, and multiply this by the central `ridge' of the spectrum, to give a 2D spectrum representing the QSO and host galaxy only. This is then subtracted from the observed 2D spectrum, to leave only the extended emission from the nebula (where present), which is shown in the right panels of Fig 1. There is visible evidence of HzLAN, with the Lyman-$\alpha$ emission extending up to several arcsec outside the galaxy spectrum, in five cases (Q1658+575, Q0758+120, Q0805+046, Q2222+051 and probably SDSSJ2228+0110).
 In this section we examine in turn the extracted 1D spectra of both the galaxies/AGN and nebulae. We fit the Lyman-$\alpha$ lines with a Gaussian on a continuum-subtracted  (by fitting high-order spline function) spectrum and, as there are no strong and narrow emission lines (e.g. HeII1640) within our relatively narrow wavelength ranges, use the QSO Lyman-$\alpha$  wavelength to define the velocity centre of each galaxy.

\subsection{Q1354+258}
This is the lowest redshift galaxy in our sample and the Lyman-$\alpha$ emission line (we fit $\lambda=3652.97\pm0.56\rm \AA$ giving $z=2.0058\pm0.0005$) is at the far blue end of the spectrograph's range where the sensitivity falls off. Even so the line is strongly detected; we measure an integrated flux  $8.32\pm 0.29\times 10^{-15}$ ergs $\rm s^{-1}$. We could not see extended line emission, but with the relatively low sensitivity at this near-UV wavelength the uncertainty on this  is large, $\sigma(F_{extended})\sim 6\times 10^{-17}$ ergs $\rm cm^{-2}s^{-1}arcsec^{-2}$.  In comparison the mean surface brightness of the extended emission detected by H91, $\simeq 1.0\times 10^{-16}$ ergs $\rm cm^{-2}s^{-1}arcsec^{-2}$. Our non-detection might suggest our choice of slit PA  was not optimal for this one source (H91 found the most extended emission to the south).

The spectrum of the galaxy/QSO (Fig 2) shows no less than four strong (observer-frame equivalent width $\rm  EW>2 \AA$) absorption lines within the broad Lyman-$\alpha$ emission line. These could be HI structures or shells associated with the QSO host galaxy, but distinct from each other as they are spread over $\sim 2000$ km $\rm s^{-1}$ in velocity, with the first two blueshifted relative to the peak emission. We detect intrinsic absorption lines for the NV(1238,1242) and SiIV(1393,1402) doublets, which all match the redshift of the second HI absorber.  CIV(1548,1550) is redward of our spectrograph range, but this QSO was also observed in the SDSS Baryon Oscillation Spectroscopic Survey (BOSS) (with identifier SDSS J135706.53+253724.4) and the SDSS spectrum (covering a redder range) shows a CIV absorption doublet within the broad CIV emission,  again with about the same redshift and $\Delta(v)$ (-444 km $\rm s^{-1}$) as the second HI absorber. 

 There are other absorption lines which do not match the wavelengths of any expected intrinsic lines, and are probably lower redshift galaxies on the line of sight (using the doublet $\lambda$ ratios we identity one MgII and two CIV absorbers); these are outside the scope of this paper and not discussed further. See Section 6 for tabulated wavelength and other data for identified  intrinsic and line-of-sight absorbers. 
  \begin{figure}
 \includegraphics[width=0.72\hsize,angle=-90]{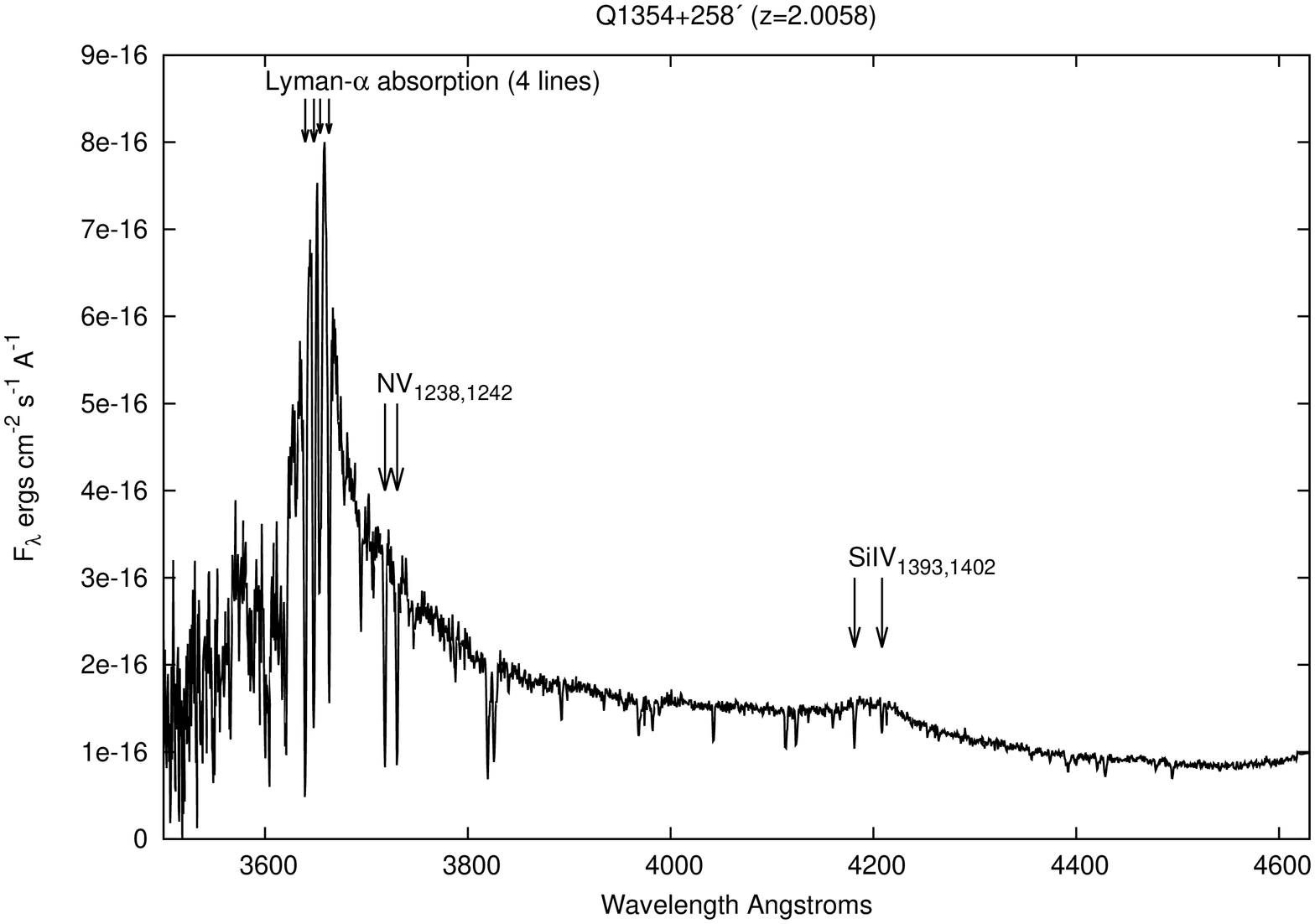}
\caption{GTC-OSIRIS spectrum of the galaxy/QSO Q1354+258 with intrinsic absorption lines labelled.} 
 \end{figure}

 \subsection{Q1658+575}
The two observing blocks for this galaxy were flagged as `attempted', due to presence of some clouds, but this seemed to have affected seriously only the first exposure (observing log: `primer espectro muy debil'), and combining the remaining 5  produced an apparently good quality spectrum (Fig 3).
The galaxy/QSO shows  a strong Lyman-$\alpha$ emission line, we fit  $\lambda=3856.04\pm 0.11\rm \AA$, giving $z=2.1728\pm 0.0001$, a flux $6.09\pm0.06\times 10^{ -15}$ ergs $\rm cm^{-2}s^{-1}$ and a FWHM $17.96\pm 0.25\rm\AA=1396\pm 18$ km $\rm s^{-1}$.

There is broad emission at $4138\rm\AA$ from SiII1304, and at $4440\rm\AA$ from SiIV, consistent with the Lyman-$\alpha$ redshift.  Blueward are many Lyman-$\alpha$ forest  lines, of which one is close enough ($\Delta(v)\simeq 1100$ km $\rm s^{-1}$) to the Lyman-$\alpha$ peak it could be an intrinsic HI absorber, and there is a stronger absorption line at about the same $\Delta(v)$  redward. 
 \begin{figure}
 \includegraphics[width=0.72\hsize,angle=-90]{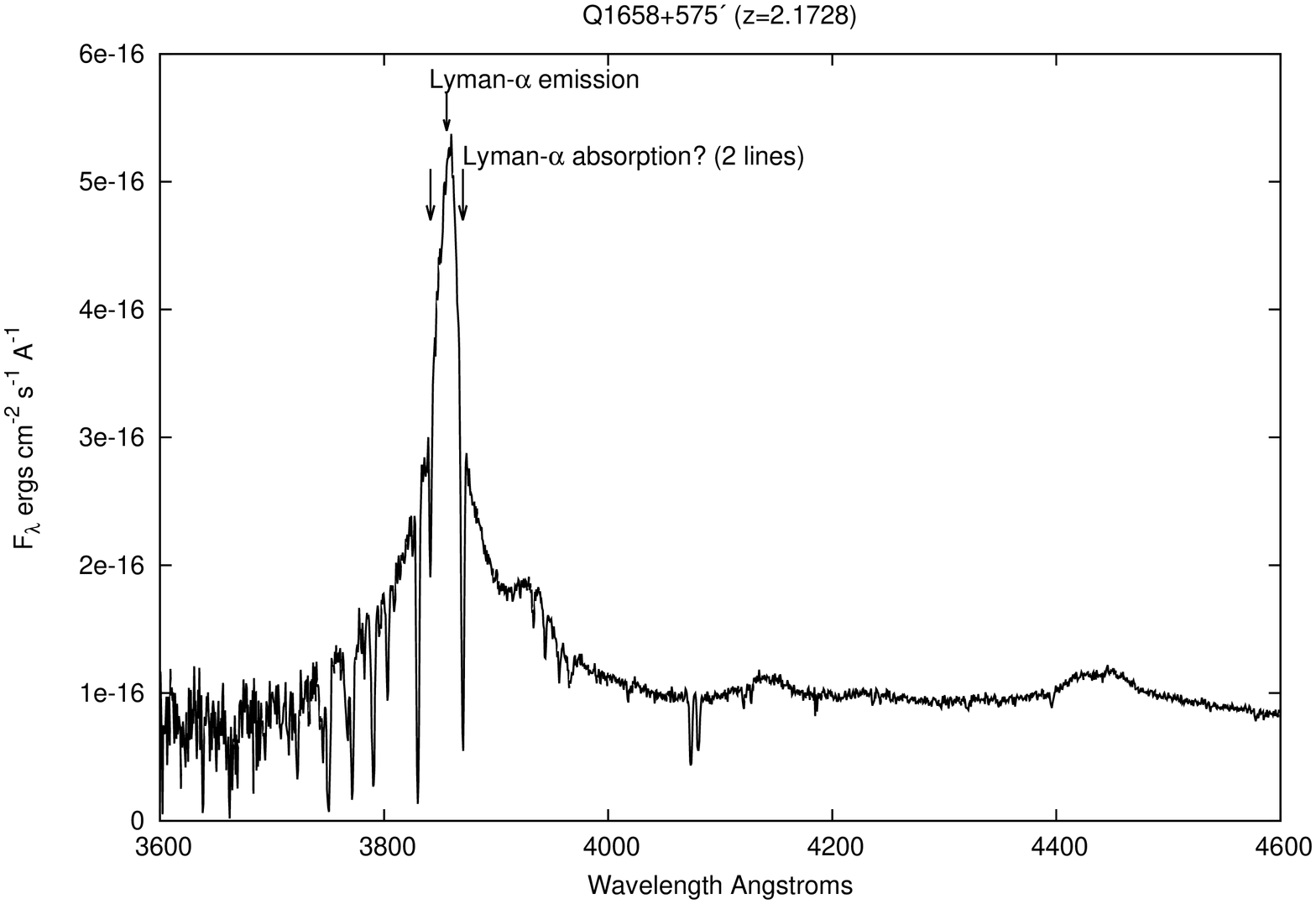}
\caption{GTC-OSIRIS spectrum of the galaxy/QSO Q1658+575.} 
 \end{figure}

On the 2D spectrum, the Lyman-$\alpha$ emission visibly extends on the left (NNE) side of the AGN, but only for 1--2 arcsec (9--17 kpc), in agreement with the WFPC2 imaging of Lehnert et al. (1999) which showed the brightest nebular emission extending in a broad plume 1--2 arcsec to the NE (it then curves to the north). This is sufficient to extract a spectrum for the HzLAN (with minimal contamination from the AGN) in a separate, parallel, narrow (4 pixels wide = 1 arcsec) aperture. 

Figure 4 compares the nebula and galaxy/QSO spectra in the region of Lyman-$\alpha$. For the nebula we see emission only in Lyman-$\alpha$, and fit this with  
$\lambda=3859.50\pm0.36\rm\AA$ (redshifted by $\Delta(v)=+269\pm 29$ km $\rm s^{-1}$     relative to the AGN emission line), $\rm FWHM=12.0\pm0.9\AA(932\pm 69$ km $\rm s^{-1}$, narrower than the AGN line), and flux  $1.92\pm0.09\times 10^{-16}$ ergs $\rm cm^{-2}s^{-1}$. 

A question of interest is whether the two absorption lines in the blue and red wings of the AGN Lyman-$\alpha$ emission line also absorb the nebular emission, as this would confirm the absorber as both intrinsic and extended (i.e. bigger than the nebula; $\geq 16$ kpc).  Indeed on our nebula spectrum there is some sign of an absorption line (we fit $3870.56\pm 0.16\rm\AA$ with EW $1.81\pm 0.46\rm\AA$) at the same wavelength as the line in the QSO spectrum, but this remains inconclusive from these data due to the  narrower emission profile of the nebula.

 \begin{figure}
 \includegraphics[width=0.72\hsize,angle=-90]{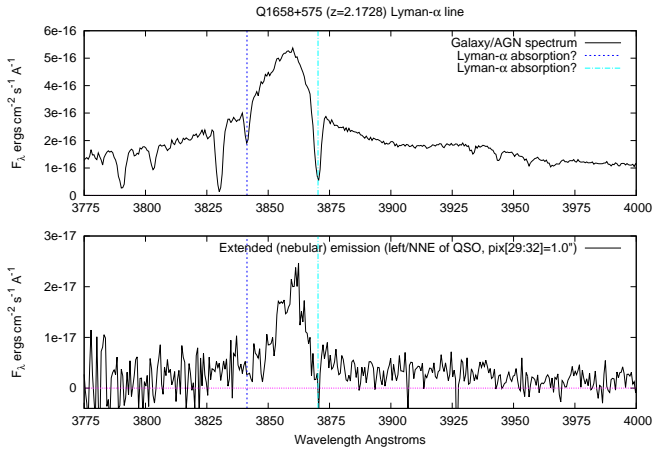}
 \caption{Comparison of the spectrum of Q1658+575 (above) and the associated nebula (extracted in an adjacent aperture on the left/NNE side) (below) in the neighbourhood of the Lyman-$\alpha$ line.}
 \end{figure}

\subsection{Q0758+120}
The galaxy/AGN spectrum (Fig 5) shows a Lyman-$\alpha$ line with a sharp peak; we fit  $\lambda=4472.95\pm 0.28\rm\AA$, giving $z=2.6804\pm 0.0002$, a flux  $4.76\pm 0.15\times 10^{-16}$ ergs $\rm cm^{-2}s^{-1}$ and a FWHM of $17.41\pm0.63 \rm \AA= 1167\pm 42$ km $\rm s^{-1}$. There are many `forest'  absorption lines in the blue wing of Lyman-$\alpha$ and one is close enough to the peak it could be intrinsic. There is broad emission from CII1334 at $4906\rm \AA$ and SiIV at $5126\rm\AA$, giving the same redshift $z=2.68$.      
\begin{figure}
 \includegraphics[width=0.72\hsize,angle=-90]{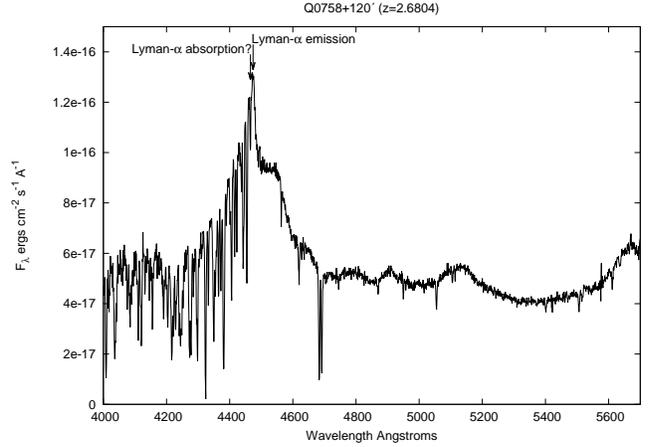}
\caption{GTC-OSIRIS spectrum of the galaxy/QSO Q0758+120.} 
 \end{figure}
 
 On the 2D spectrum, extended Lyman-$\alpha$ emission is visible, almost all on the left side (SE), where it extends to some 26 pixels= 6.6 arcsec (54 kpc) from the nucleus (H91 also found this HzLAN to be large). A spectrum of the nebula was extracted, isolated from the AGN's,  in a separate aperture of width 16 pixels = 4.0 arcsec (there is no obvious velocity shift across this distance). This spectrum (Fig 6) shows emission only in Lyman-$\alpha$  (HeII1640 will be out of the spectrograph range); we fit with $\lambda=4476.84\pm 0.40\rm \AA$, a flux $1.56\pm0.11\times 10^{-16}$ ergs $\rm cm^{-2}s^{-1}$, and FWHM $11.59\pm1.20 \rm \AA=776\pm81$ km $\rm s^{-1}$. The nebular emission line is again less broad than the AGN's, and redshifted by $\Delta(v)=+261\pm 33$ km $\rm s^{-1}$.
 
This QSO was previously observed spectroscopically (at lower resolution) by Heckman et al. (1991b), who found the Q0758+120 nebula spectrum dominated by the Lyman-$\alpha$ line with
 $F_{Ly\alpha}=4\times 10^{-16}$  ergs $\rm cm^{-2}s^{-1}$ (through slit), $z=2.682\pm0.003$, $\rm FWHM=1100\pm400$ km $\rm s^{-1}$, and $\Delta(v)=-100\pm 400$ km $\rm s^{-1}$ relative to the QSO.

 \begin{figure}
 \includegraphics[width=0.72\hsize,angle=-90]{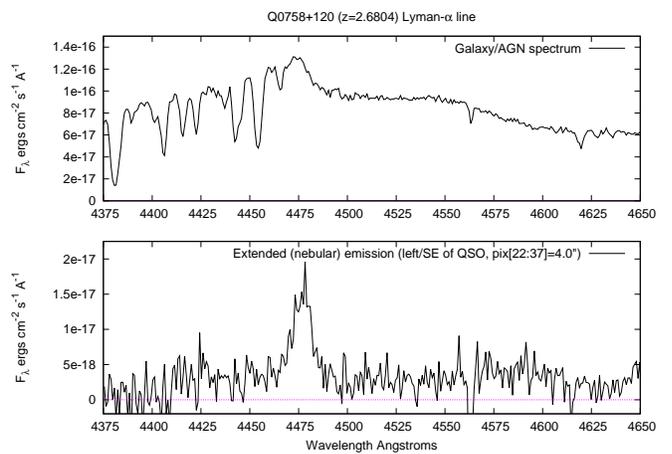}
 \caption{Comparison of the spectrum of Q0758+120 (above) and the associated nebula (extracted in an adjacent aperture on the left/SE side) (below) in the neighbourhood of the Lyman-$\alpha$ line.}
 \end{figure}

\subsection{Q0805+046}
The spectrum (Fig 7) is dominated by strong Lyman-$\alpha$ emission; we fit $\lambda=4711.63\pm 0.11\rm\AA$ giving $z=2.8768\pm0.0001$, flux $1.339\pm0.004\times 10^{-14}$ ergs $\rm cm^{-2} s^{-1}$ (the brightest of any QSO in this sample) and a very broad FWHM $50.43\pm0.17\rm\AA=3209\pm11$ km $\rm s^{-1}$. 
 The most obvious feature is a huge absorption line neatly bisecting the quasar's Lyman-$\alpha$. 
Blueward of this, there are many `Lyman-$\alpha$ forest' absorption lines, redward,  one absorption doublet is identified as intrinsic NV, others as metal lines from lower redshift galaxies. A broad emission feature at $\lambda=5060\rm\AA$ could be SiII1304 at the QSO redshift $z\simeq 2.88$.
   \begin{figure}
 \includegraphics[width=0.72\hsize,angle=-90]{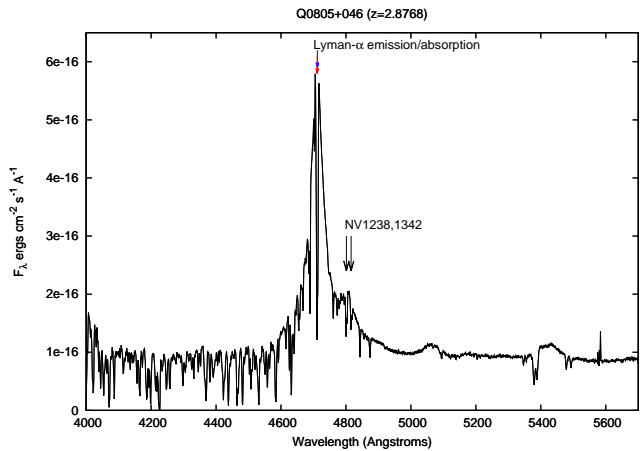}
\caption{GTC-OSIRIS spectrum of the galaxy/QSO Q0805+046.} 
 \end{figure}

The 2D spectrum shows extended Lyman-$\alpha$ emission, visible on both sides of the QSO, and spanning as much as $\sim9$ arcsec (72 kpc), with greater brightness and extension on the left (SE) side. H91 found this to be the largest and brightest nebula in their sample, similarly with `the bulk of the high surface brightness emission arising from the SE side of the QSO)'. We extract the nebular spectrum in two strips, one of width 13 pixels (3.3 arcsec) on the left, the other a 10 pixel aperture (2.5 arcsec) on the right/NW. Both (Fig 8) show a single Lyman-$\alpha$ emission line, at about the same wavelengths as the half of the QSO line redwards of the bisecting absorption. For the  SE side nebula we fit $\lambda=4718.81\pm0.71\rm\AA$, flux $3.87\pm0.37\times 10^{-16}$ ergs $\rm cm^{-2} s^{-1}$ and FWHM $17.02\pm1.89\rm \AA=1059\pm118$ km $\rm s^{-1}$, and for the NW side line, $\lambda=4717.11\pm0.99\rm\AA$, flux $1.43\pm0.17\times 10^{-16}$ ergs $\rm cm^{-2} s^{-1}$ and FWHM $15.96\pm2.62\rm\AA=1014\pm 167$ km $\rm s^{-1}$.

 Thus again the nebular line, on both sides, is less broad than, and redshifted with respect to, the QSO line; with $\Delta(v)=+457\pm46$ (SE) and $+349\pm 63$ km $\rm s^{-1}$ (NW). The blue edge of the nebular emission is near the wavelength of the strong absorption line, but on Figure 8 there is still some nebular flux at the line centre, so the absorber likely has a smaller covering factor for the nebula than for  the AGN. 
   \begin{figure*}
 \includegraphics[width=0.72\hsize,angle=-90]{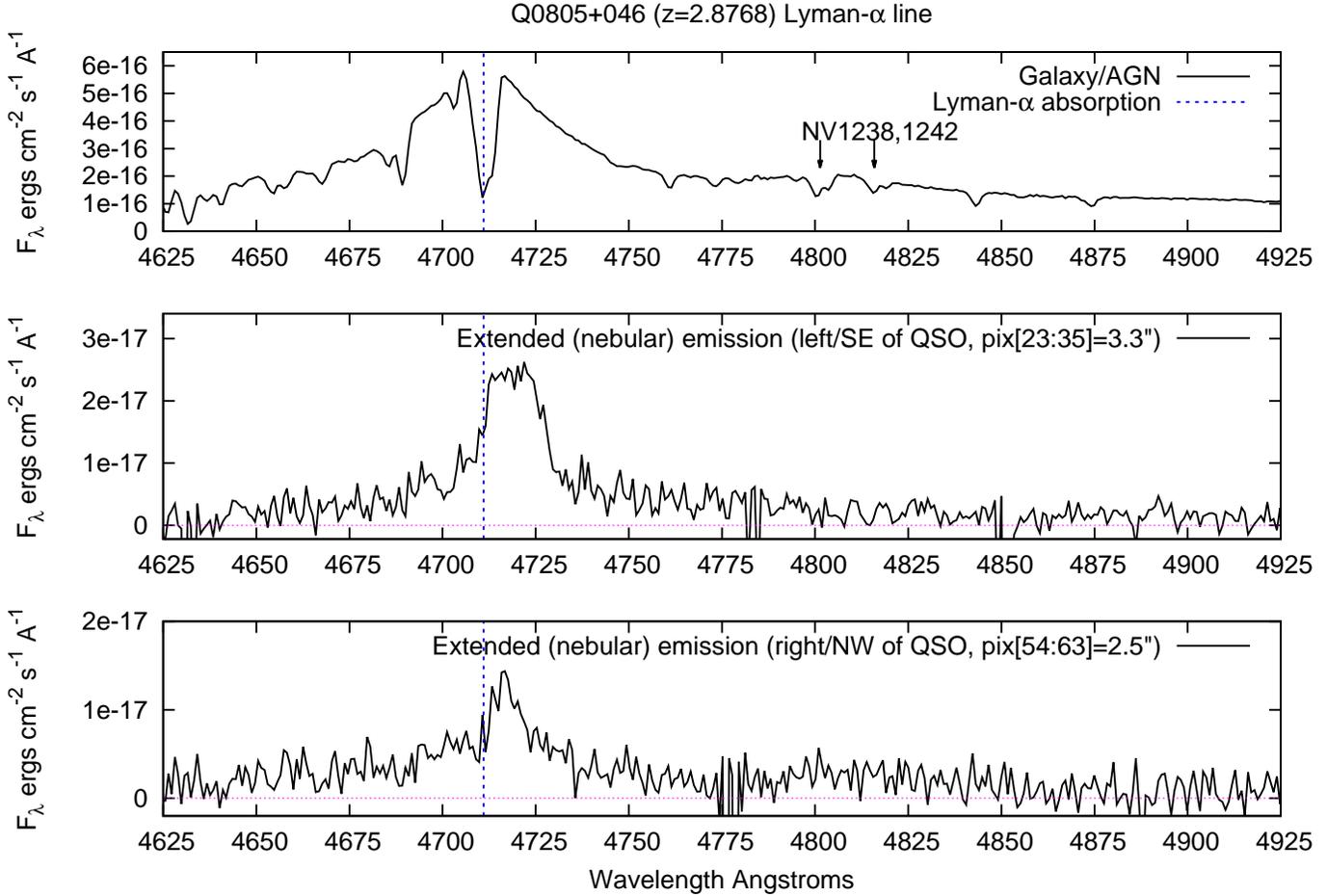}
\caption{Comparison of the spectrum of Q0805+046 (above) and the associated nebula, extracted in adjacent apertures on the left/SE (mid) and right/NW (below) sides, shown in the neighbourhood of the Lyman-$\alpha$ line.}
 \end{figure*}

 Heckman et al. (1991b) obtained a spectrum for this nebula and measured a Lyman-$\alpha$ flux $2.5\times  10^{-15}$ ergs $\rm cm^{-2}s^{-1}$ , $\rm FWHM=1100\pm200$ and a velocity offset $\Delta(v)=200\pm200$ km $\rm s^{-1}$.  They also detect a weak (7\% strength of Lyman-$\alpha$) narrow CIV emission line, which cannot be verified here as it falls redward of our spectrograph range. For this and the previous QSO, the through-slit fluxes of Heckman et al. (1991b) are $\sim 0.5$ dex higher than ours, simply because they sample larger areas of the nebulae, due to their wider slit (1.5 compared to 0.8 arcsec) and our cautious (i.e. well away from AGN) choice of apertures. Their kinematic measurements and ours are consistent within the uncertainties.

\subsection{Q2338+042} 
This observing block was flagged as `attempted' due to unspecified bad conditions and indeed  all our spectra appear very faint. We combine the 3 exposures to obtain the spectrum on Fig 9. The flux and signal-to-noise are low compared to all the other H91 targets, suggesting there was significant cloud. Even so, the Lyman-$\alpha$ emission line is clearly visible, split in half by a very strong (saturated) absorption line. For the emission we fit $\lambda=4365.86\pm 0.43\rm\AA$, giving $z=2.5923\pm 0.0004$, an integrated flux $2.42\pm0.07\times 10^{-16}$ ergs $\rm cm^{-2}s^{-1}$ (only 2\% the previous object!) and  FWHM $44.29\pm 2.19\rm\AA=3042\pm 150$ km $\rm s^{-1}$.
The other feature of note  is strong and broad CIV emission, again bisected by a strong absorption line (probably from the same absorber) in which saturation has merged the two components of the 1548,1550 doublet.   

On the 2D image we cannot see extended emission and, due to poor signal-to-noise and very non-photometric conditions, cannot even give a lower limit on this. Extended Lyman-$\alpha$ emission was previously detected here by H91, with a NW/SE orientation,  and again by Lehnert and Becker (1998) in Keck spectroscopy and by Lehnert et al. (1999) in HST narrowband imaging. On this basis our  slit angle ($50^{\circ}$ W of N) was correct and the non-detection can probably be blamed on poor conditions.
    \begin{figure}
 \includegraphics[width=0.72\hsize,angle=-90]{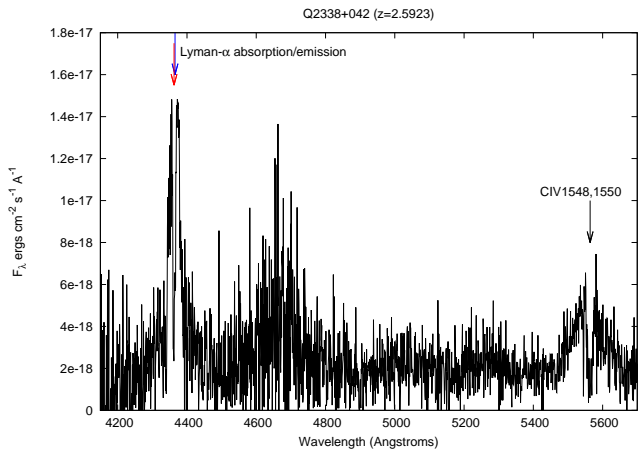}
\caption{GTC-OSIRIS spectrum of the galaxy/QSO Q2338+042.} 
 \end{figure}

 \subsection{Q2222+051}
 The galaxy/QSO spectrum (Fig 10) shows strong Lyman-$\alpha$ emission -- we fit $\lambda=4038.56\pm 0.13\rm\AA$, giving  $z=2.3230\pm 0.0001$, flux $2.51\pm0.06\times 10^{-15}$  ergs $\rm cm^{-2} s^{-1}$ and FWHM 
 $11.78\pm 0.33\rm\AA= 874\pm 22$ km $\rm s^{-1}$.  The CIV emission peak is also very prominent -- we fit $\lambda=5142.82\pm 0.37$ ($z\simeq 2.320$) and a flux $2.52\pm0.06\times 10^{-15}$ --  giving a high CIV/Ly$\alpha$ ratio of $\simeq 1.0$, and the continuum is redder than the other QSOs (despite our correction for the strong Galactic reddening which had reduced the flux by $\sim 45\%$ at the blue end). 
 
 We identify one single strong absorption line as probably intrinsic SiII1260. The absorption line in the blue wing of the Lyman-$\alpha$ emission, together with the strong NV and CIV doublets, are likely intrinsic, but very blueshifted $\Delta(v)\sim 1000$ km $\rm s^{-1}$.
 It is also possible that the blueshifted absorption is produced by a neighbouring galaxy, such as the large disk galaxy `4C 05.84 ER1' observed by Stockton et al. (2008) using NICMOS, $\sim 20$ arcsec W of the QSO and (on the basis of photometry) at the same redshift, with a massive ($M_*\simeq 3.3\times 10^{11}\rm M_{\odot}$) and old ($\sim 1.02$ Gyr) stellar population.

\begin{figure}
 \includegraphics[width=0.72\hsize,angle=-90]{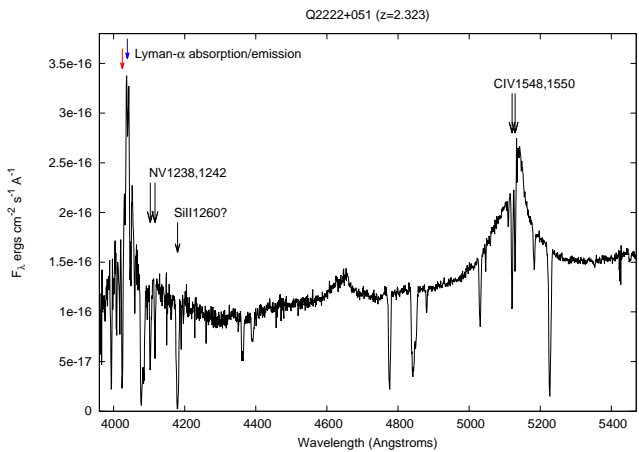}
\caption{GTC-OSIRIS spectrum of the galaxy/QSO Q2222+051.} 
 \end{figure}
 On the 2D spectrum, extended emission is clearly visible, almost all on the right (SW or SSW) of the AGN spectrum, where is extends to at least 5 arcsec (42 kpc) from the nucleus (H91 similarly reported `the brightest and most extended emission to the SSW'). There is also at least a hint of much fainter emission extending for about 1 arcsec on the left (NE), but the signal-to-noise was too poor to extract a useful spectrum here. Hence 
we extract the nebular spectrum on the right of the AGN, in a 12 pixel  (3.0 arcsec) aperture (Fig 11). It shows Lyman-$\alpha$, with $\lambda=4037.97\pm0.23\rm \AA$, FWHM of $5.53\pm0.68\rm\AA=411\pm 50$ km $\rm s^{-1}$ and a flux $6.80\pm 0.61\times 10^{-16}$  ergs $\rm cm^{-2} s^{-1}$. Again the nebula Lyman-$\alpha$ line is narrower than the QSO's, but in this case it is not redshifted with respect to it, with 
 $\Delta(v)=-44\pm 20$ km $\rm s^{-1}$.  In this aperture we also find a much fainter emission line 
 at $\lambda=5449.20\pm0.18 \rm \AA$, matching closely the expected wavelength for HeII1640,  for the Lyman-$\alpha$ redshift of either the QSO (formally, with $\Delta(v)=-23\pm16$ km $\rm s^{-1}$) or the nebula.
 In the 3.0 arcsec aperture we measure the HeII1640 flux as  $3.69\pm 0.58\times 10^{-17}$  ergs $\rm cm^{-2} s^{-1}$, giving the Lyman-$\alpha$/HeII ratio here as $18.4\pm4$, in the same range as $22\pm 4$ for the nebula of TXS1436+157 (Humphrey et al. 2013) and $12.73\pm 0.23$ for the radio galaxy 4C40.36 (Sanchez and Humphrey 2009). These ratios are all consistent with AGN photoionization as the primary source of Lyman-$\alpha$, with or without a smaller contribution from star-formation.
 The HeII line is very narrow with $\rm FWHM= 2.57\pm 0.61 \rm\AA = 141\pm 35$ km $\rm s^{-1}$, shows no significant velocity gradient within the 3 arcsec aperture,  and  is not seen at all outside of this aperture, not even in the QSO spectrum. HeII emission might be associated only with the brightest radio hotspot, 2--3 arcsec SW of the nucleus.  
  \begin{figure}
 \includegraphics[width=0.72\hsize,angle=-90]{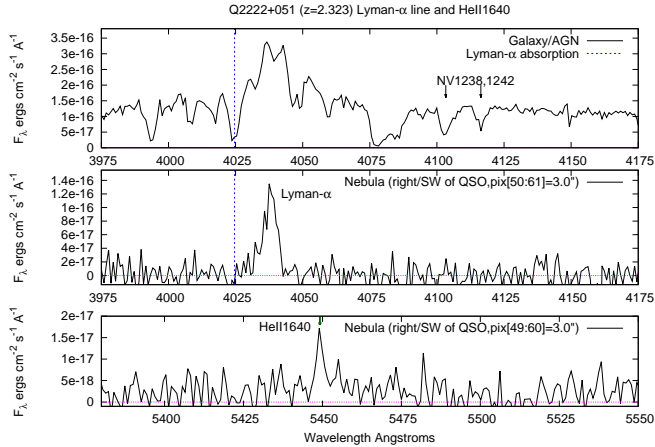}
 \caption{Comparison of the spectrum of Q2222+0151 (above) and the associated nebula on the right/SW side (mid) in the neighbourhood of the Lyman-$\alpha$ emission and absorption lines, and (below) the faint and narrow HeII1640 line seen for the nebula only.}
 \end{figure}
\subsection{SDSSJ2228+0110}
We also observed the much fainter $z\sim 6$ QSO, SDSSJ2228+0110, using the R2500I grating. Inevitably, the signal-to-noise is poorer than for the H91 sources, furthermore the Lyman-$\alpha$ line is redshifted into a region of the near-infrared with a high sky background and many sky lines, adding to the noise. The spectrum (Fig 11) shows broad  Lyman-$\alpha$ emission in the midst of narrower bands of noise from the sky, with the continuum visible redwards of the line. Immediately blueward, there is a sharp break to a near-zero flux, as expected (due to absorption from the intergalactic medium) for all sources at this high redshift (e.g. Roche et al. 2012).

  For the Lyman-$\alpha$ line we fit $\lambda=8389.51\pm0.29\rm\AA$ giving $z=5.9030\pm 0.0002$, a flux $2.86\pm0.04\times 10^{ -16}$ ergs $\rm cm^{-2}s^{-1}$ and FWHM $33.27\pm 0.68\rm \AA=1189\pm 24$ km $\rm s^{-1}$. The peak of the line appears to be split by a possible absorption line with  $\lambda=8385.14\pm0.09\rm \AA$ and $EW=1.20\pm0.15\rm\AA$. If genuine it would be slightly blueshifted -- $\Delta(v)=-156\pm15$ km $\rm s^{-1}$. Two brighter objects happened to fall on our spectrograph slit and neither of their spectra show an absorption feature at this wavelength, implying it is intrinsic to the QSO, but sky lines increase the noise here and more data is needed to confirm this.  There may be weaker broad emission at NV1238,1242.
   Zeimann et al. (2011), from their Keck LRIS spectrum, measure a   slightly broader FWHM 1890 km $\rm s^{-1}$ for the Lyman-$\alpha$, and a rest-frame (i.e. divided by $1+z$) EW of $21.9\rm \AA$; we agree, finding $22.1\rm\AA$ ( they do not mention any absorption line). 
 \begin{figure}
 \includegraphics[width=0.72\hsize,angle=-90]{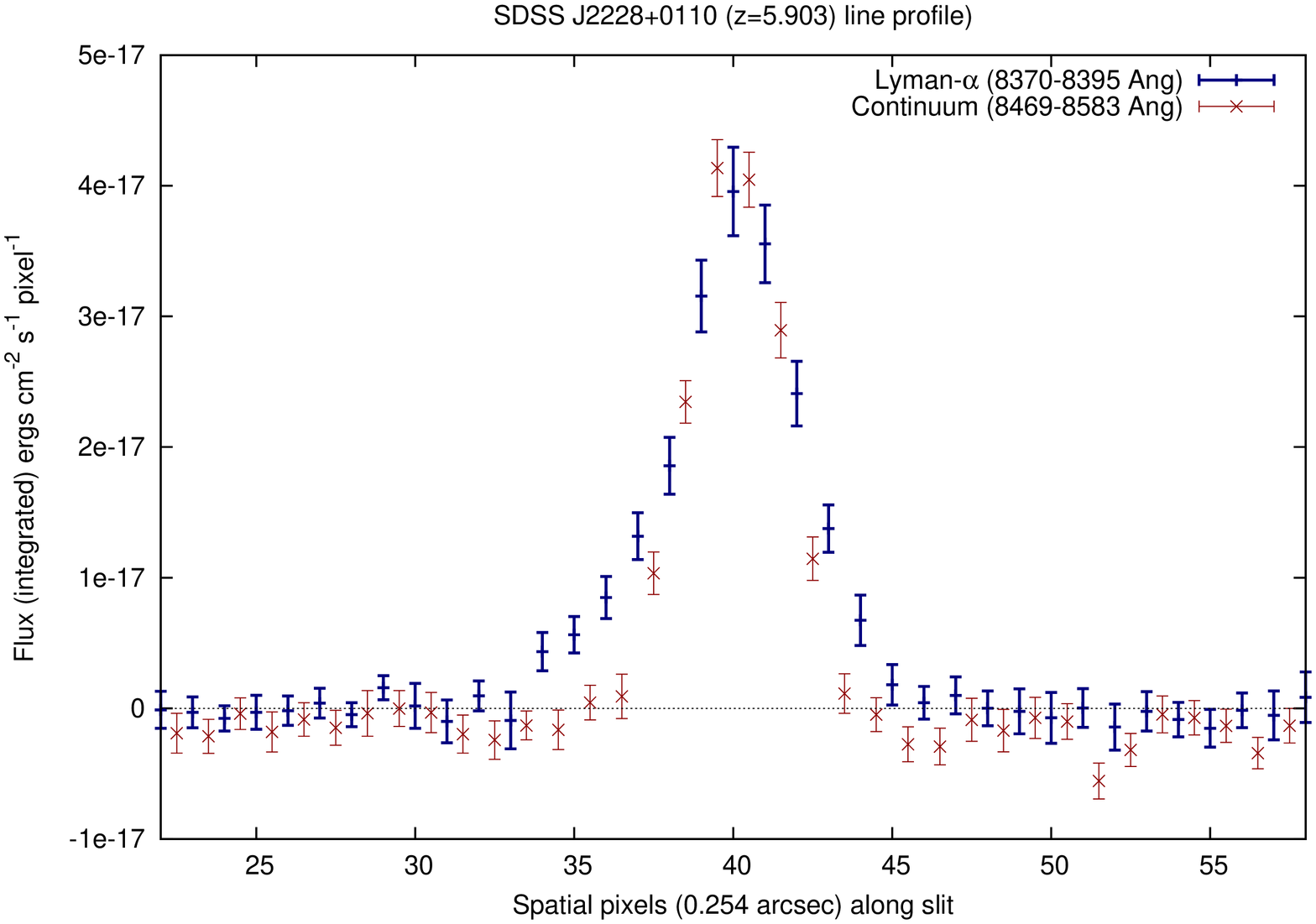}
 \caption{Spatial profile of the R2500I spectrum of  SDSSJ2228+0110 summed over the Lyman-$\alpha$ line and compared the profile summed over a nearby region of continuum. The Lyman-$\alpha$ emission is more extended (and asymmetric), implying the QSO is surrounded by a HzLAN of at least modest ($\sim 9$ kpc) size.}
 \includegraphics[width=0.72\hsize,angle=-90]{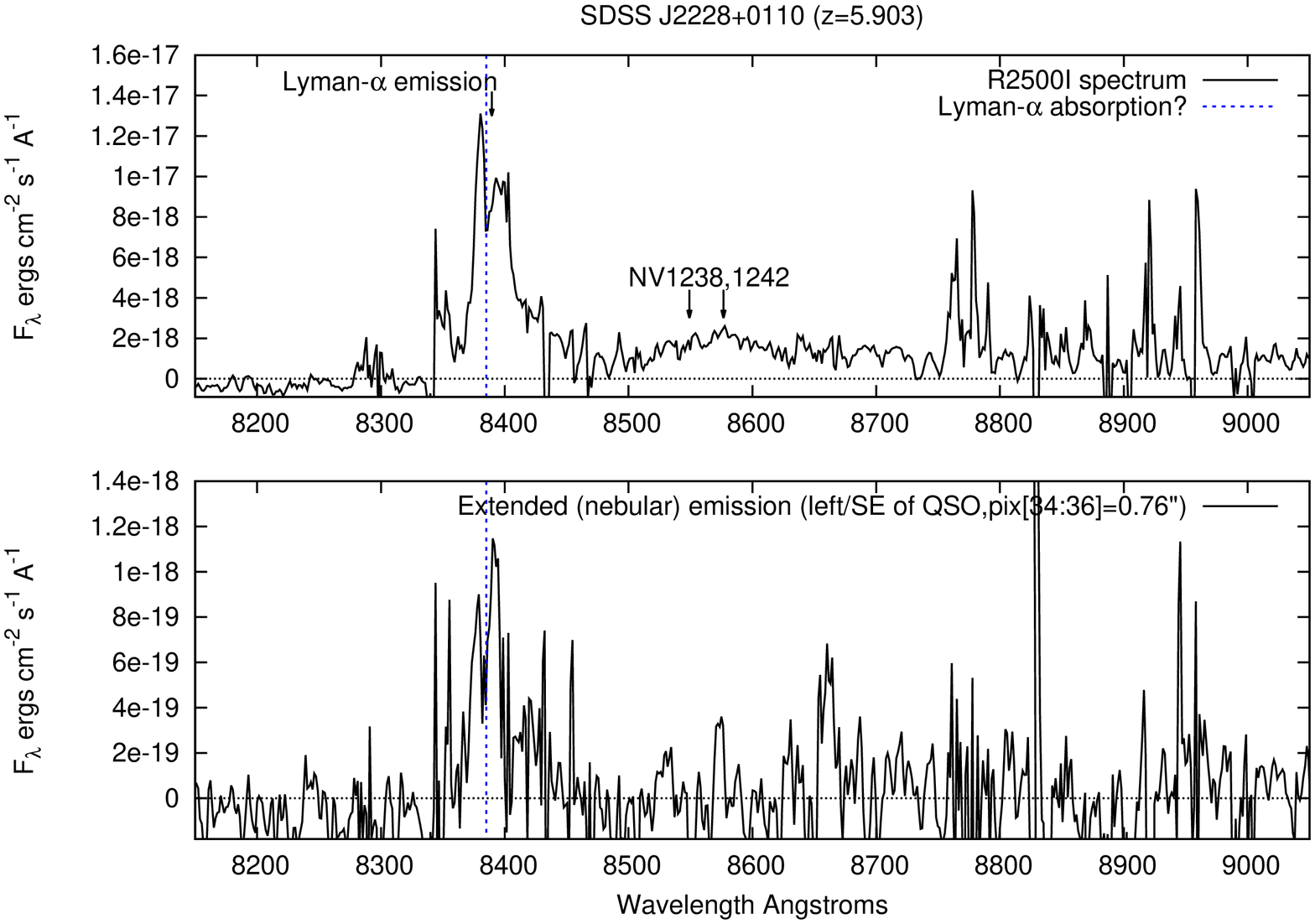}
 \caption{R2500I spectrum of  SDSSJ2228+0110 (above) and the associated nebula on the left/SE side (below), showing only the region of interest in the neighbourhood of the Lyman-$\alpha$ line and break. These two spectra (unlike others in this paper) have been slightly smoothed before plotting by convolution with a FWHM=2 pixel Gaussian.}
 \end{figure}

 On our 2D spectrum there is some sign that Lyman-$\alpha$  emission is a little ($\sim 1.5$ arcsec) extended to the left of the galaxy/QSOs spectrum. So (Fig 12) we compare the spatial profiles of the spectra summed over most of the the Lyman-$\alpha$ line and over a region of the continuum with a similar total flux a short distance redward (in both cases avoiding spectral regions with strong sky lines). The spatial profile is more extended on the Lyman-$\alpha$ line than for  the continuum, with  a Gaussian-fit FWHM of $1.19\pm 0.04$ compared to $0.86\pm 0.02$ arcsec, and it also looks  asymmetric. This is evidence  of an associated HzLAN, and we extract a nebula spectrum in a narrow 3 pixel (0.76 arcsec) aperture just leftward (SE) of the AGN (Fig 13); it shows  noisy but significant Lyman-$\alpha$ emission (amidst a continuum consistent with zero). We measure a flux $2.02\pm0.46\times 10^{-17}$ ergs $\rm cm^{-2}s^{-1}$ and centre $\lambda=8386.26\pm 2.41\rm\AA$; the FWHM is poorly constrained with $27.5\pm 13.5\rm\AA$ ($983\pm 483$ km $\rm s^{-1}$), and the velocity shift would be $\Delta(v)= -116\pm 87$,  consistent with zero.
 
 Thus we find evidence that this much higher redshift RL-QSO (at Universe age 1.0 Gyr) shows a bright, extended (at least 1.5 arcsec = 9 kpc) emission-line nebula, and possibly blueshifted absorption, similar to its many $2<z<3$ counterparts e.g. in the H91 sample. But at $z\sim 6$, this is one of very few examples of HzLAN: see Section 5.2.
 
  \section{Spatially Resolved Kinematics of the Nebulae}

   \begin{figure}

\includegraphics[width=0.72\hsize,angle=-90]{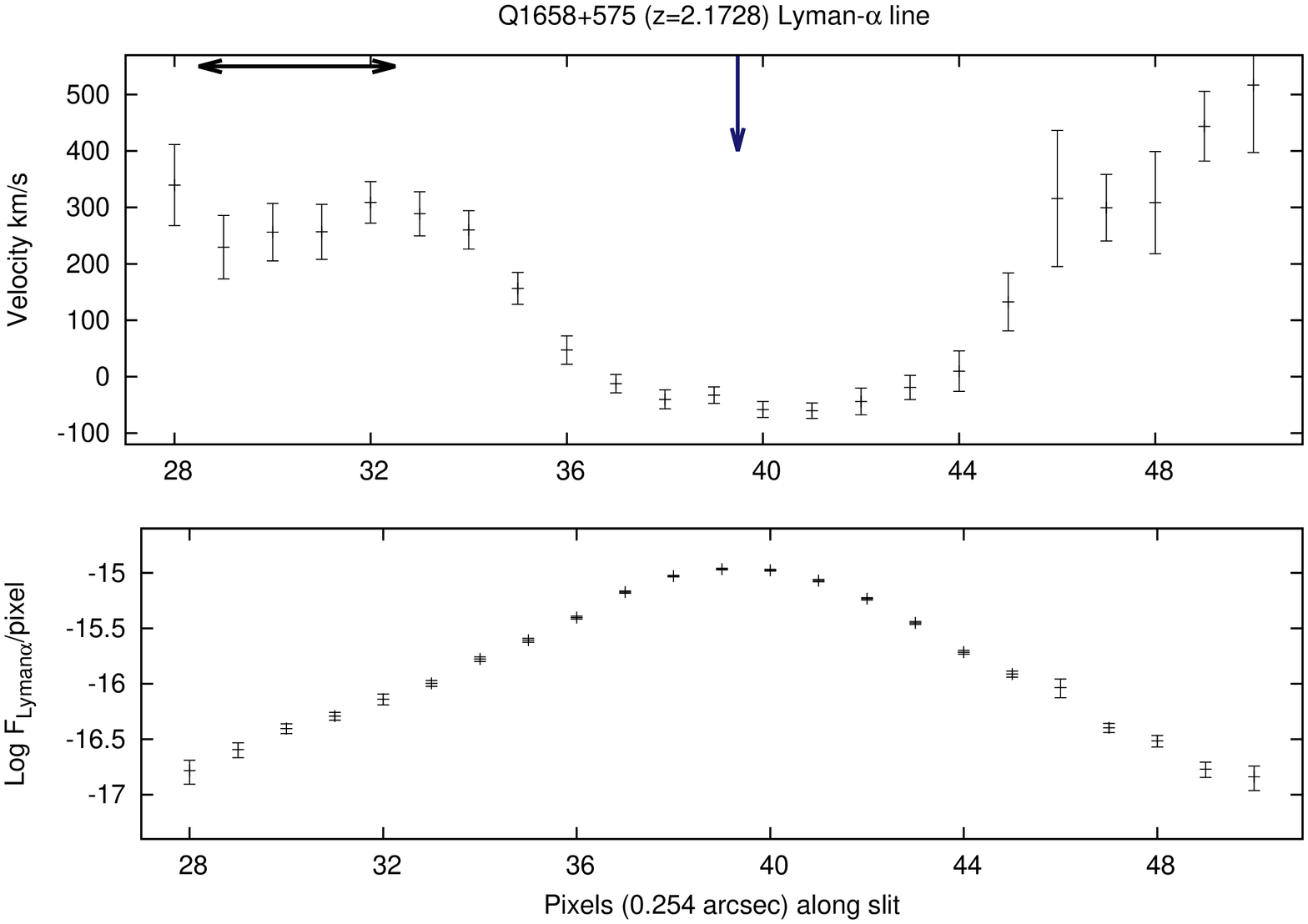} 
 
 \includegraphics[width=0.72\hsize,angle=-90]{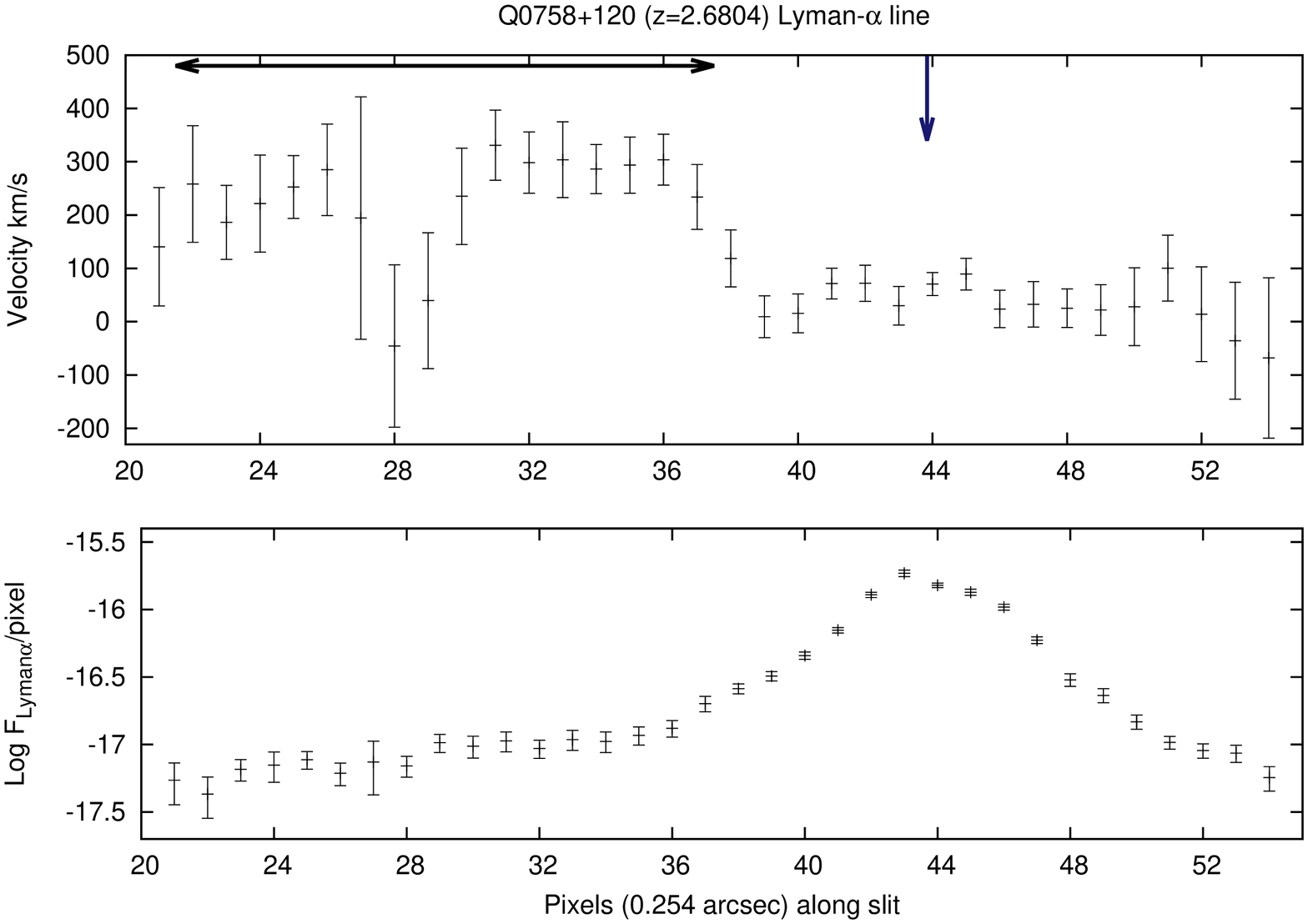}
 \includegraphics[width=0.72\hsize,angle=-90]{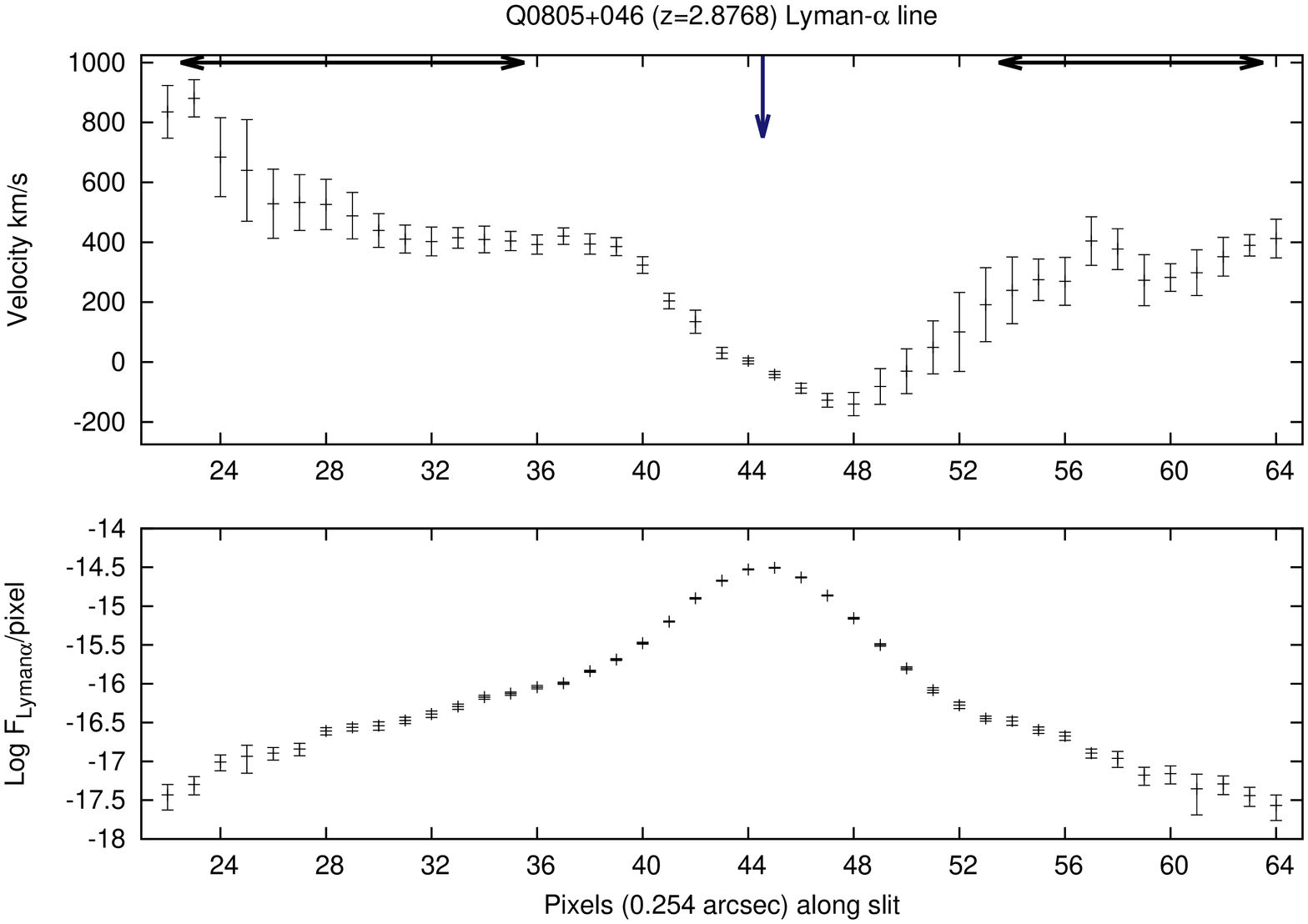}


\includegraphics[width=0.72\hsize,angle=-90]{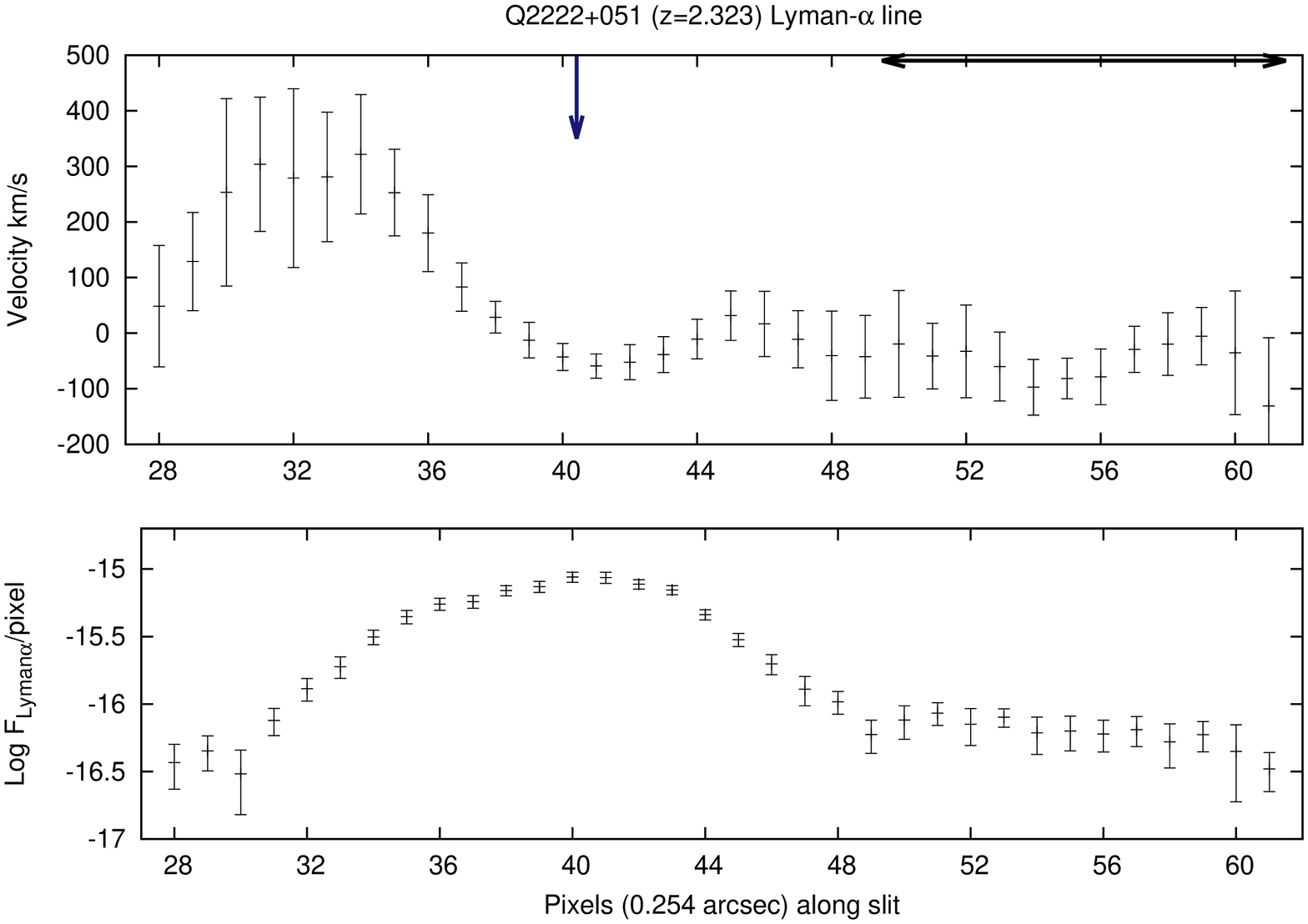}
\end{figure}
\begin{figure}
\includegraphics[width=0.72\hsize,angle=-90]{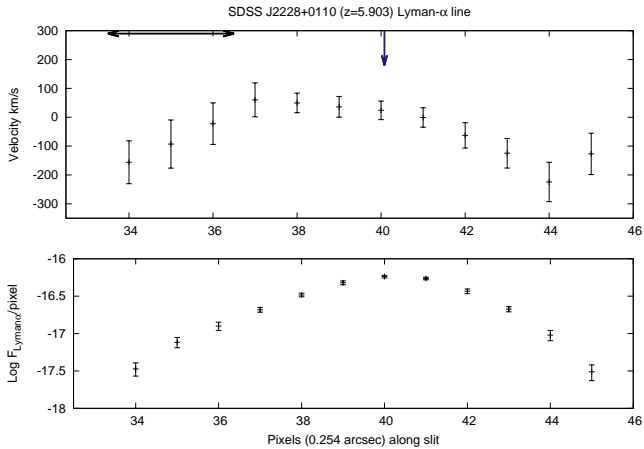}
 \caption{Spatially resolved Lyman-$\alpha$ kinematics from the 5 of our 2D spectra where extended emission is apparent on one or both sides of the central AGN (marked by arrow). Pairs of plots show (above) Lyman-$\alpha$ emission line velocity, derived from the fitted wavelength relative to that in the AGN spectrum, $\Delta(v) = c(\lambda - \lambda_{AGN} )/\lambda_{AGN}$ . The doubleheaded horizontal arrows show the apertures in which the nebula spectra were extracted (in Section 3).The vertical arrows show the centre of the QSO spectrum.  }
 \end{figure}

  In the previous Section we found that three of the nebulae are redshifted by 250--460 km $\rm s^{-1}$ relative to the AGN. We can make further use of the high resolution in the two available (spatial and spectral) dimensions to study the {\it internal kinematics} of the nebulae, from the wavelength of peak Lyman-$\alpha$ emission as a function of distance along the slit. 
 
 For each flux-calibrated 2D spectrum, we proceed along the slit in the spatial direction, pixel-by-pixel, and interactively fit a Gaussian to the continuum-subtracted Lyman-$\alpha$ line in each 1-pixel strip, wherever it is visible above the sky noise.  Before the fitting, each 2D spectrum is smoothed by a $3\times3$ pixel boxcar, sufficient to reduce sky noise by a factor $\sim 3$ and thus aid the profile fitting but with very minor loss of resolution. These fits give us the Lyman-$\alpha$ central wavelength as a function of spatial distance, across the nebula, host galaxy and AGN. Figure 14 shows the Lyman-$\alpha$ velocity offset in the one-pixel strips, calculated as $\Delta(v)=c (\lambda-\lambda_{AGN})/\lambda_{AGN}$ (where $\lambda_{AGN}$ is the Lyman-$\alpha$ wavelength measured for the QSO in the previous section),  and (below) line flux (per pixel).

  We see again that three of the  emission nebulae (Q0805+046, Q1658+575 and Q0758+120)  are redshifted by $\Delta(v)\sim 200$--500 km $\rm s^{-1}$ and the other two are approximately at rest on the line-of-sight axis, relative to the QSO. Examination of the spectra at higher spatial resolution has not revealed large monotonic velocity gradients (i.e. several hundred km $\rm s^{-1}$) across the nebulae that might  indicate kinematics dominated by rotation (at any large angle to the line of sight). The V-shaped velocity curves of Q1658+575 and Q0805+046 flatten in the outer parts and so appear to be due to a systematic redshifting of the whole nebula relative to the QSO, rather than to the nebula's internal motion.

In Q0805+046, we find a small velocity shift, $\sim 100$   km $\rm s^{-1}$ over $\sim 8$ arcsec  (64 kpc), across the nebula (see also Fig 8), and
a much steeper velocity gradient, in the same direction, across the central peak of the profile (pixels 40--48); $\sim 400$ km $\rm s^{-1}$ across 2 arcsec (16 kpc). The strong absorption line at $4711\rm\AA$ may complicate interpretation and velocity measurements, but a close-up examination of the line profile (Fig 15) indicates a genuine shift in the emission line; note the highest emission peak lies blueward of the bisecting absorption line on the left of the AGN, but is redward on the right side. One possibility for producing a steep gradient might be Lyman-$\alpha$ emission from star-formation in a rapidly rotating host galaxy, or an ongoing merger, but the line is very broad here ($\rm FWHM\sim 3000$ km $\rm s^{-1}$) and we do not see an obvious narrow component. More likely, the velocity gradient might be attributed to the nebular emission, if its central region is very bright and had much more active/turbulent kinematics than the relatively quiescent outer region, with a rapidly rotating inner disk and/or a localised outflow (e.g. in a narrow jet towards the observer, as in Nesvadba et al. 2008).

   The faint left/NE side of the Q2222+051 nebula also appears on our plot to be redshifted (by as much as $\sim 200$ km $\rm s^{-1}$)  relative to the QSO and (non-redshifted) right side, but the signal-to-noise is very poor here. There is also a peak in the FWHM. If this is genuine it is probably due to a localised outflow, rather than an overall rotation, because this velocity curve remains very flat at $\Delta(v)\simeq 0$ on the right/SW side.

   \begin{figure}
\includegraphics[width=0.72\hsize,angle=-90]{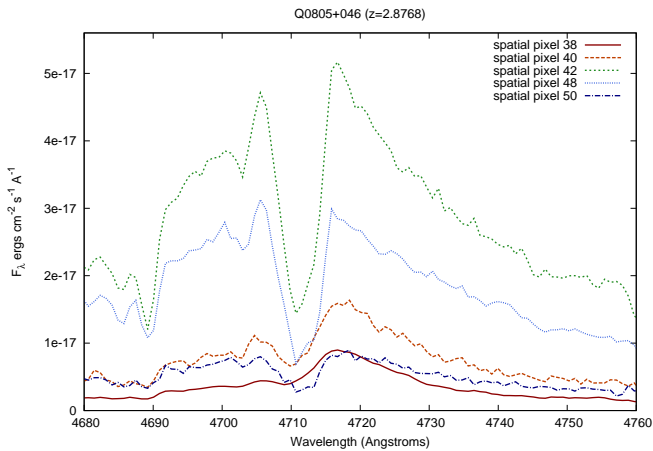}
 \caption{Spectrum of the Q0805+046 Lyman-$\alpha$ line in a series of one-pixel strips crossing the centre of the spectrum (QSO and host galaxy, at pixel 44.5 here). There is a subtle but definite blueward shift in the broad line, which on the left side of the QSO (i.e. pixels 38--42) has higher flux to the red of the bisecting absorption line, but on the right of the QSO, at pixel 48, it is actually higher on the blueward side.}
 \end{figure}

 \section{Discussion}
 \subsection{Kinematics of the Giant Nebulae}
 \begin{table}
\begin{tabular}{lccc}
\hline
Nebula & Ly$\alpha$ SB ergs & $\Delta(v)$ & FWHM \\
                     & $\rm cm^{-2}s^{-1}arcsec^{-2}$ & km $\rm s^{-1}$ & km $\rm s^{-1}$  \\
  \hline                   
 Q1658+575 &    $1.90\times 10^{-16}$ & $+269\pm 29$ & $932\pm 69$ \\
 Q0758+120 &    $0.48\times 10^{-16}$ & $+261\pm 33$ & $776\pm 81$ \\
 Q0805+046(SE) & $1.47\times 10^{-16}$ & $+457\pm 46$ & $1059\pm 81$ \\
 Q0805+046(NW) & $0.70\times 10^{-16}$ & $+349\pm 63$ & $1014\pm 167$ \\
 Q2222+051 & $2.79\times 10^{-16}$ & $-44\pm 20$ & $411\pm 50$ \\
 J2228+0110 & $0.44\times 10^{-16}$ & $-116\pm 87$ & $983\pm 483$ \\
 \hline
 \end{tabular}
 \caption{Summary of the properties of the Lyman-$\alpha$ emitting nebulae, as measured in strips parallel to the AGN spectra (Section 3). To enable a comparison the fluxes are converted into mean Lyman-$\alpha$ surface brightness (SB) by dividing by (strip width $\times$ slit width).}
  \end{table}

  We examined the nebular Lyman-$\alpha$ lines (summarized in Table 2) and found little indication of systematic velocity gradients across the nebulae. This implies that if these nebulae are rotating structures, the rotation axes are close to the line-of-sight. However, 3/5 of the nebulae do show overall redshifts in their Lyman-$\alpha$ emission relative to the AGN, of $\sim 250$--460 km $\rm s^{-1}$.  
 For Q2338+042 we were unsuccessful in seeing a nebula due to poor conditions; but previously  Lehnert and Becker (1998) detected the extended nebula up to 5 arcsec on both sides of the AGN, and with Keck LRIS data and essentially the same methods as ours investigated its kinematics. From their findings the Q2338+042 nebula fits the same pattern as the rest of our sample with a broad FWHM of up to 1300 km $\rm s^{-1}$, a small redshift relative to the AGN ($\Delta(v)\sim 100$--200 km $\rm s^{-1}$) and little or no internal velocity gradient.
 
  Previously, Villar-Mart\'in et al. (2003), performed long-slit Keck and VLT spectroscopy of 10 radio galaxies with extensive nebulae ($\sim 10$--14 arcsec) at $2<z<3$.  The HeII1640 and Lyman-$\alpha$ emission lines gave similar $\Delta(v)$ curves, showing velocity shifts across the nebula diameters ranging from near-zero to $\simeq 600$ km 
  $\rm s^{-1}$. In at least 4 and probably 6 of the 10 RGs the velocity fields were consistent with rotating systems, with the velocities giving dynamical masses a few $\times 10^{12}\rm M_{\odot}$  (consistent with massive galaxies). Orbiting satellite galaxies might contribute to the velocity shifts, but it was considered the line emission would probably be dominated by the gaseous halo. 
 
 Of course the observed $\Delta(v)$ would also depend (as sin $i$) on the orientation of the rotation axis relative to the line of sight. 
 Hence if (i) radio galaxies and RL-QSOs differ only by orientation,  with the axis or `bi-cone' where the AGN emission is not obscured by the torus being close to the line-of-sight in the RL-QSOs but closer to the sky plane in HzRGs, and (ii) the nebulae are rotating and/or undergoing ordered, anisotropic motion (infall or outflow)  with an axis aligned with the optical/radio emission axis of the AGN, then the nebula velocity maps (as observed on the line-of-sight) would show  different patterns for radio galaxies and RL-QSOs. Less orientation-dependent  nebular properties (luminosity, line ratios, diameter, velocity FWHM etc.) would remain closely similar. Indeed there is already evidence for this.
 
Using VLT-VIMOS integral field spectroscopy, Villar-Mart\'in et al. (2006 and 2007) studied the Lyman-$\alpha$ nebula kinematics of four further $z\sim2.5$ radio galaxies. Three had double components in Lyman-$\alpha$ (matching the radio peaks) with velocity separation a few 100 km $\rm s^{-1}$, suggesting rotation  and/or two orbiting or merging galaxies. The other nebula (MRC 1558-003) is rounded with the centre redshifted by $\sim 400$ km $\rm s^{-1}$ relative to the  outermost regions, a pattern  more suggestive of infall. 

Humphrey et al. (2007), using a sample of 11 HzRG nebulae at $2.3<z<3.6$, which had internal velocity shifts $\Delta(v)= 200$--770 km $\rm s^{-1}$, found a correlation of asymmetries: in addition to the known radio/Lyman-$\alpha$ alignment, the side of the HzLAN brighter in both 
 Lyman-$\alpha$ and radio (i.e. the {\it near} side) was almost always the more redshifted side. They interpreted this as evidence for (anisotropic) infall of the nebulae onto the host galaxies (ultimately to fuel the AGN and star-formation). The greater prominence of infall in the nebula velocity map of MRC 1558-003 could then be explained by its orientation, with its emission cone axis closer to the line of sight  ($\sim 20^{\circ}$ from) than for the others; there is evidence for this in the spectrum of this galaxy which shows  broad  $\rm H\alpha$ emission from the AGN broad line region. As this galaxy is closer to the RL-QSO class than most radio galaxies (if only by its alignment), the implication is the nebulae of RL-QSOs would show similar (i.e. infall-dominated) velocity maps.
 
 Christensen et al. (2006) observed a sample of high-$z$ QSOs with the Potsdam Multi Aperture Spectrophotometer (PMAS) integral field spectrograph, and for two detected HzLAN of sufficient Lyman-$\alpha$ brightness (similar to our sample) to map the line velocity. One (the radio-loud Q1759+7539) showed a fairly uniform $\Delta(v)\simeq +300$ km $\rm s^{-1}$ relative to the QSO emission-line, the other (the radio-quiet Q1425+606) a radial gradient with $\Delta(v)\simeq 600$ decreasing to 200 km $\rm s^{-1}$ at 3.5 arcsec (which is similar to MRC 1558-003 mentioned above). Most recently, Humphrey et al (2013) in GTC-OSIRIS spectroscopy of the $z=2.54$ RL-QSO TXS 1436+157 found a relatively small, 110 km $\rm s^{-1}$ redward shift in HeII1640 from the AGN to the nebula (on its east side): this again was interpreted as evidence for infall of the nebula, at $\sim 10$--$\rm 100M_{\odot}yr^{-1}$ onto the galaxy.

In our RL-QSO sample,  the nebular Lyman-$\alpha$ have a mean FWHM  of 863 km $\rm s^{-1}$ with range  400--1100 km $\rm s^{-1}$, which is similar to radio galaxy nebulae; a little less than the FWHM of 700--1600 km $\rm s^{-1}$ in van Ojik et al. (1997), a little greater than in the  Villar-Mart\'in et al. (2003) sample with 250--850 km $\rm s^{-1}$, and in the same 400--1500 km $\rm s^{-1}$ range as the Villar-Mart\'in et al. (2007) integral-field spectroscopy of 3 HzRGs.
 Similarly Christensen et al. (2006) found Lyman-$\alpha$ FWHM$\simeq 500$ km  $\rm s^{-1}$ for the bright nebulae around two QSOs.  
   However, whereas almost all of the radio galaxy nebulae  in Villar-Mart\'in et al. (2003, 2006, 2007) and Humphrey et al. (2007) show substantial velocity gradients (shears), amounting to $\Delta(v)=200$--770 km $\rm s^{-1}$ across the nebular diameter, our 5 RL-QSO nebulae have much flatter velocity profiles (Fig 14). Neither do we find examples of nebulae with more redshifted centres (like Q1425+606 and MRC 1558-003), although such an effect (if limited to a small central area) might have been hidden within the much stronger broad-line emission from the QSOs. These RL-QSOs appear more similar  to Q1759+7539 of Christensen et al. (2006), a RL-QSO at $z\simeq 3$ with a large nebula redshifted by 200--300 km $\rm s^{-1}$ with little or no internal velocity gradient.

      In the Q0805+046 nebula, the brighter side appears to be slightly ($\Delta(v)\simeq 100$ km $\rm s^{-1}$) more redshifted, in agreement with the correlation described by Humphrey et al (2007) for HzRGs. We also find a much steeper ($\sim 400$ km $\rm s^{-1}$ in 2 arcsec) velocity gradient across the centre or host galaxy of Q0805+046, suggesting more complex kinematics at the very centre of the nebula, such as fast rotation or a localised outflow. On the  basis of the interesting velocity curve and high surface brightness nebula, this object especially would merit further investigation by high-resolution integral-field spectroscopy.  Q2222+051 is possibly a counterexample to the Humphrey at al. (2007) correlation, in that our line fitting found a region on the fainter, left side to be the more redshifted (the line is also very broad in FWHM here) -- this might even be an outflow directed away from the observer.

 It should be noted that a Lyman-$\alpha$ emission line can be redshifted (by up to a few 100 km $\rm s^{-1}$) in a number of ways, e.g. infall of an emission nebula, an outflowing jet directed away from the observer, or thirdly, as in the models of Verhamme et al (2008) and Schaerer et al. (2011), which represent galaxies as central starbursts within expanding HI shells. In these models the approaching HI shell causes a blueshifted Lyman-$\alpha$ absorption line while backscattering from the receding side redshifts the Lyman-$\alpha$ emission that reaches the observer. However, this may be less effective for the much broader emission lines of powerful AGN. Detection of polarisation in a few HzLAN (e.g. Hayes et al. 2011; Humphrey et al. 2013b) implies that some percentage of their Lyman-$\alpha$ may indeed be back-scattered, yet we consider it more likely that for HzLAN the majority of the observed photons come from the ionized nebulae directly and show its true velocity, rather than being the nuclear photons reflected in a moving HI `mirror'. We did find  some blueshifted intrinsic absorbers in the QSO spectra, but their lines are narrow and unsaturated (except in Q2338+042), suggesting the effects of HI scattering on Lyman-$\alpha$ are relatively small. Also, detection of extended HeII emission tracing the same velocity as Lyman-$\alpha$, from the Q2222+051 nebula here, and many others (e.g. Villar-Mart\'in et al. 2003; S\'anchez and Humphrey 2009), suggests the photoionized gas is in HzLAN observed directly, as HI  could not back-scatter the HeII line in the same way as Lyman-$\alpha$. And the correlation of Humphrey et al. (2007), followed by at least Q0805+046 in this sample, implies that redshifted Lyman-$\alpha$ usually comes (directly) from the near side of the nebulae.
 
 Martin et al (2014) discovered that a $z=2.8$ QSO with extended Lyman-$\alpha$ emission was also
 associated with two long Lyman-$\alpha$ emitting filaments, with relative $\Delta(v)$ +100 and +600 km$\rm s^{-1}$, which were believed to be accreting onto the host galaxy. 
 Recent models predict (Dekel et al. 2008, S\'anchez Almeida et al. 2014) that massive galaxies at $z>2$ (in general) are being fuelled by accretion from the `cosmic web'  at rates $\sim 100$--200 $\rm M_{\odot} yr^{-1}$.  This web of filaments forms HzLAN around (within $\sim 100$ kpc of) active galaxies, where the accreting gas becomes orders of magnitude more luminous due to fluorescence (Cantalupo et al. 2014). The filaments, extending hundreds of kpc (so far exceeding the sizes of any HI-rich galaxies)  must be observed directly in Lyman-$\alpha$, giving their true kinematics, and the same might apply for the HzLAN, which are like bright knots in the web. 
 
These inflowing streams could be simultaneous with powerful AGN/starburst-driven outflows, most likely to be seen as blueshifted emission regions with high FWHM and very steep velocity gradients (e.g. Nesvadba et al. 2008;  Villar-Mart\'in et al. 2011).  As noted above, such signatures of outflows might be seen in one or two locations in our data, but most of the extended Lyman-$\alpha$ in this sample is more consistent with nebulae slowly infalling (e.g. as in Weidinger et al. 2005), or that appear stationary (e.g. Q2222+051 and the $z\sim2.7$ `blob' of  Yang et al. 2014) with $\Delta(v)\sim0$. The latter might actually have mixed infall and outflow, or infall with scattering as in the models of Dijkstra, Haiman and Spaans (2005), although their lines appear single rather than double-peaked, which could set limits on this.
  
 For the Q0805+046 nebula  the (14 arcsec) visible  diameter and integrated Lyman-$\alpha$ flux ($10^{-14.1}$ ergs $\rm cm^{-2}s^{-1}$) from H91 give radius $r=56$ kpc and $L_{\rm Ly\alpha}=10^{44.77}$ ergs $\rm s^{-1}$ (the highest in this sample). We measure a high gas velocity $\Delta(v)$= +457 km $\rm s^{-1}$, and from this could estimate the total mass within $r$  as between $v^2 r/2G$ (assuming infalling gas is in free fall) and $v^2 r/G$ (for a circular orbit), giving 1.36--$2.72\times 10^{12}\rm M_{\odot}$. If the gas infalls to a nucleus at $\sim 1$ kpc (taking 118 Myr if velocity remains constant) and the density profile inside the nebula is $\propto r^{-2}$ (giving a $r^{-1}$ force of gravity) we estimate that to power all the Lyman-$\alpha$ emission by the gravitational energy difference would require a (very high)  infall rate of $\sim 1800~\rm M_{\odot} yr^{-1}$. The lower infall rates expected from models would not be sufficient to account for the
  $L_{\rm Ly\alpha}$, e.g. in the detailed cosmological-hydrodynamical simulations of Goerdt et al. (2010), Lyman-$\alpha$ nebulae powered almost entirely by the gravitational potential energy from instreaming gas onto massive haloes  would be an order of magnitude less luminous than these HzLAN. 
 
In contrast, Humphrey et al. (2007) proposed that powerful HzLAN like this  would have moderate but substantial infall rates onto the host galaxy of  $\sim 200~\rm M_{\odot} yr^{-1}$. For their assumed volume-averaged HII density $\sim 100$ atoms $\rm cm^{-3}$ and volume filling factor $10^{-5}$, $m_H=1.6727\times 10^{-24}$ g and a factor 1.35 to include the mass of helium etc.,  the Q0805+046 nebula within  visible radius 56 kpc would have a total gas mass $2.45\times 10^{10}\rm M_{\odot}$. Dividing this by our estimated infall time gives an infall rate 208 $\rm M_{\odot} yr^{-1}$. Such  infall rates, in good agreement with the above-mentioned models with accretion from streams, would certainly be sufficient to power the Lyman-$\alpha$ emission if the principal photoionization source is the AGN (almost certainly the case here) and a significant fraction ($\geq 10\%$) of the accreting gas goes to feed the SMBH.

 As discussed above the differences, and the similarities, in the nebula spectra of RL-QSOs and HzRG are most likely due to the similar natures but different orientation of these AGN.  Thus in  the more on-axis (small angle $i$) RL-QSOs, systematic velocity gradients across the nebulae may be suppressed on the line-of-sight direction by $\sim \rm sin$ $i$ factors, without this having much influence on the observed surface brightness, diameter and line FWHM of the nebulae. In addition, this orientation could have brought components of systematic infall (expected for models of accretion) in the emission nebula kinematics into greater prominence in the QSO long-slit spectra. This we may have seen in three of the nebulae, from their $\Delta(v)=260$--450 km $\rm s^{-1}$. The other two nebulae, with $\Delta(v)\sim 0$, might be on average at rest on this axis.

 \subsection{Discovery of a ${\bf z\sim 6}$ QSO Nebula}
  Our second major finding from the OSIRIS spectroscopy is an extended ($\geq 10$ kpc) Lyman-$\alpha$ emission nebula around the $z=5.9$ QSO SDSSJ2228+0110, which is a new discovery for this galaxy. There are many small Lyman-$\alpha$ emitting galaxies at $z\sim 6$ but extended nebulae are much rarer.  Indeed this is the highest redshift detection of a HzLAN associated with a radio-loud AGN. However, there is one other known example of a HzLAN associated with a QSO at these redshifts, CFHQSJ2329-0301 at $z=6.43$, first detected by Goto et al. (2009) in narrow $z$-band imaging, and confirmed by spectroscopy (Willott et al. 2011 and separately by Goto et al. 2012).  This nebula is similar to our example in extent and Lyman-$\alpha$ surface brightness, $\sim 15$ kpc and $\sim 4\times   10^{-17}$ ergs $\rm cm^{-2}s^{-1}arcsec^{-2}$ from Willott et al. (2011), and apparently in kinematics  in that the Lyman-$\alpha$ line showed no significant $\Delta(v)$ relative to the AGN or velocity gradients, and a similar FWHM measured as 640 (Willott et al. 2011) and $707\pm 232$ (Goto et al. 2012)  km $\rm s^{-1}$.
  
 A third example is the radio-quiet Himiko HzLAN of Ouchi (2009, 2012) at $z=6.6$, similarly luminous and also $\sim 7$ kpc in radius, but perhaps different in nature as it is believed to be a triple merging system undergoing intense star-formation.
 The Lyman-$\alpha$ surface brightness of all 3 $z\sim 6$ HzLAN, 2--$5\times 10^{-17}$ ergs $\rm cm^{-2}s^{-1}arcsec^{-2}$, is relatively low  compared to the bright nebulae of  RGs and RL-QSOs found  $2<z<3$ (Table 2, H91 etc.), but this is accounted for entirely by the higher redshift (a factor 8 dimming from $z=2.5$ to 6), and the only obvious intrinsic difference is the smaller spatial extent of the $z\sim 6$ nebulae which might be related to  less massive host galaxies at these `cosmic dawn' epochs.

\subsection{Intrinsic Absorption Systems}
   \begin{table}
\begin{tabular}{lcccc}
Wavelength & EW($\rm \AA$)  & ID & Redshift & $\Delta(v)$ \\ 
\hline
 $3639.66\pm0.06$  & $2.30\pm0.13$ & Ly$\alpha$& 1.9949 & -1092\\
$3648.22\pm0.06$  & $2.49\pm0.10$ &   Ly$\alpha$ & 2.0019 & -390\\
   $3654.31\pm0.08$ & $2.28\pm0.10$  &   Ly$\alpha$ & 2.0069 & +110\\
 $3663.10\pm0.09$  & $2.09\pm0.12$ &    Ly$\alpha$ & 2.0141 &  +831\\
  $3718.27\pm0.14$ & $1.66\pm 0.20$ & NV1238 & 2.0024 & -344\\
   $3730.29\pm0.15$  & $1.24\pm 0.21$ & NV1242 & 2.0024 & -339\\
 $4181.35\pm0.10$  & $0.62\pm0.06$ & SiIV1393 & 2.0009 & -486\\
  $4208.36\pm 0.12$  & $0.36\pm 0.06$ & SiIV1402 & 2.0009 & -488 \\
     $4646.60_{(SDSS)} $ & 1.915 & CIV1548 & 2.0013 & -444 \\
\medskip    
   $4654.33_{(SDSS)}$  & 2.218 & CIV1550 &  2.0013 & -444 \\

$3819.72\pm 0.13$ & $3.15\pm0.40$ & CIV1548 & 1.4690&\\
  $3826.06\pm0.13$ & $2.57\pm0.33$ & CIV1550 & 1.4679&\\
$4113.55\pm0.10$ & $0.71\pm0.07$ & MgII2796 & 0.4714&\\  
  $4123.95\pm0.15$ & $0.52\pm0.07$ & MgII2803 & 0.4714&\\
   $4420.79\pm 0.41$ & $0.23\pm 0.11$ & CIV1548 & 1.8562&\\
 $4428.64\pm 0.32$ & $0.53\pm 0.12$ & CIV1550 &  1.8566&\\
 \hline 
 \end{tabular}
 \caption{Absorption lines we identify in our GTC spectrum of Q1354+258 (plus the CIV lines from the SDSS-BOSS spectrum) - the lines (above) intrinsic to the galaxy, with the velocity offset $\Delta(v)$ in  km $\rm s^{-1}$ derived from the difference in redshift relative to the peak of Lyman-$\alpha$ emission (below)  from lower redshift galaxies on the line of sight}.
\begin{tabular}{lcccc}
Wavelength & EW($\rm \AA$)  & ID & Redshift & $\Delta(v)$ \\ 
\hline
$3841.36\pm0.07$ & $0.66\pm0.05$ & Ly$\alpha$& 2.1607 & -1142\\
\medskip
$3870.32\pm0.04$ & $2.26\pm0.06$ &  Ly$\alpha$& 2.1846 &+1110\\

$3933.32\pm 0.17$ & $0.30\pm0.06$ & MgII2796 & 0.4070&\\
$3943.85\pm0.15$ & $0.39\pm0.06$ &  MgII2803 & 0.4071&\\
$3956.46\pm0.18$ & $0.41\pm0.09$ & MgII2796 & 0.4152&\\  
$3966.04\pm0.48$ & $0.43\pm0.11$ & MgII2803 & 0.4150&\\
$4073.85\pm 0.07$ & $1.47\pm 0.08$ & CIV1548 & 1.6321&\\
$4080.47\pm 0.10$ & $1.13\pm 0.08$ & CIV1550 &  1.6320&\\
 \hline 
 \end{tabular}
 \caption{Absorption lines in GTC spectrum of Q1658+575.}
  \begin{tabular}{lcccc}
Wavelength & EW($\rm \AA$)  & ID & Redshift & $\Delta(v)$ \\ 
\hline
\medskip
$4465.62\pm 0.16$ & $0.39\pm 0.05$ & Ly$\alpha$? & 2.6744 & -491\\

$4683.63\pm0.05$ & $3.16\pm 0.10$ & CIV1548 & 2.0260&\\
$4691.45\pm 0.06$ & $2.65\pm 0.08$ & CIV1550 & 2.0260&\\
$5105.31\pm 0.45$ & $0.28\pm 0.07$ & CIV1548 & 2.2985&\\
$5113.80\pm 0.40$ & $0.13\pm 0.07$ & CIV1550 & 2.2985&\\
 \hline 
 \end{tabular}
 \caption{Absorption lines in GTC spectrum of Q0758+120.}
 \begin{tabular}{lcccc}
Wavelength & EW($\rm \AA$)  & ID & Redshift & $\Delta(v)$ \\ 
\hline
$4711.42\pm 0.01$ & $4.35\pm 0.02$ & Ly$\alpha$ & 2.8767  & -13 \\
$4801.13\pm 0.07$ & $1.55\pm 0.05$ & NV1238 & 2.8767 & -16 \\
\medskip
$4815.91\pm 0.10$ & $0.987\pm 0.04$ & NV1242 & 2.8761 & -57 \\

$4761.02\pm0.07$ & $0.53\pm0.03$ & MgII2796 & 0.7031& \\
$4773.10\pm0.11$ & $0.43\pm0.03$ & MgII2803 & 0.7030&\\
$4843.20\pm0.06$ & $0.95\pm0.04$ & SiIV1393 & 2.4759&\\
$4874.05\pm0.11$ & $0.88\pm0.05$ & SiIV1402 & 2.4756&\\
$5346.59\pm0.18$ & $0.51\pm0.05$ & CIV1548 & 2.4543&\\
$5355.31\pm0.27$ & $0.21\pm0.05$ & CIV1550 & 2.4543&\\
$5379.78\pm0.09$ & $3.89\pm0.08$ & CIV1548 & 2.4754&\\
$5387.86\pm 0.08$ & $3.39\pm0.09$ & CIV1550 & 2.4752&\\
$5478.76\pm0.15$ & $1.25\pm0.07$ & MgII2796 & 0.9598&\\
$5492.89\pm0.19$ & $0.79\pm0.06$ & MgII2803 & 0.9598&\\
 \hline 
 \end{tabular}
 \caption{Absorption lines in GTC spectrum of Q0805+046.}
 \end{table}
\begin{table}
 \begin{tabular}{lcccc}
Wavelength & EW($\rm \AA$)  & ID & Redshift & $\Delta(v)$ \\ 
\hline
$4362.68\pm 0.14$ & $7.15\pm 0.24$ & Ly$\alpha$ & 2.5887 & -218 \\
$5564.00\pm 0.36$ & $13.15\pm 0.38$ & $\rm CIV_{1548}^{1550}$ & 2.5909 & -39\\
 \hline 
 \end{tabular}
 \caption{Absorption lines in GTC spectrum of Q2338+042 (as the CIV 1548,1550 doublet is blended, the redshift is from an averaged wavelength of $1549.48\rm \AA$). }
  \begin{tabular}{lcccc}
Wavelength & EW($\rm \AA$)  & ID & Redshift & $\Delta(v)$ \\ 
\hline
$4024.62\pm 0.06$ & $3.09\pm 0.11$ & Ly$\alpha$ & 2.3112  & -1035\\
$4103.31\pm0.16$ & $2.68\pm 0.18$ & NV1238 & 2.3132 & -882\\
$4116.37\pm0.20$ & $1.24\pm 0.16$ & NV1242 & 2.3131 & -892\\
$4179.85\pm0.10$ & $7.77\pm0.23$ &  SiII1260 & 2.3171 & -525\\
$5120.83\pm0.06$ & $1.87\pm0.06$ & CIV1548 & 2.3085 & -1302\\
\medskip
$5129.33\pm0.08$ & $0.96\pm0.05$ &CIV1550 & 2.3085 & -1302\\

$4078.33\pm 0.12$ & $6.32\pm0.37$ & CIV1548 & 1.6349 &\\
$4084.75\pm0.21$ & $3.36\pm0.28$ & CIV1550 & 1.6347 &\\
$4362.58\pm 0.18$& $3.08\pm0.19$ & SiIV1393 & 2.1310 &\\
$4390.50\pm0.43$ & $1.82\pm 0.31$ & SiIV1402 &  2.1308 &\\
 \hline 
 \end{tabular}
 \caption{Absorption lines in GTC spectrum of Q2222+051}

 \begin{tabular}{lcc}
 \hline
 Galaxy & Line(s) detected & $\Delta(v)$ km $\rm s^{-1}$ \\
\hline
Q1354+258 & Ly$\alpha$ & -1092 \\
      & Ly$\alpha$,$\rm NV^{1242}_{1238}$,$\rm SiIV^{1402}_{1393}$, $\rm CIV^{1550}_{1548}$ & [-419] \\
                         & Ly$\alpha$ & +110 \\
\medskip                         
                          & Ly$\alpha$ & +831\\

Q1658+575   & Ly$\alpha$ & -1142 \\
\medskip
                          & Ly$\alpha$ & +1110\\
\medskip
Q0758+120 & Ly$\alpha$? & -491 \\

\medskip
Q0805+046  & Ly$\alpha$, $\rm NV^{1242}_{1238}$& [-29] \\

\medskip
Q2338+042 & Ly$\alpha$, $\rm CIV^{1550}_{1548}$ & $[-128]$ \\

Q2222+051 & Ly$\alpha$, $\rm NV^{1242}_{1238}$,$\rm CIV^{1550}_{1548}$ & $[-1082]$ \\
\medskip
                       & SiII1260 & -525\\

J2228+0110 &   Ly$\alpha$? & -156 \\
\hline
\end{tabular}
\caption{Summary of (probable) intrinsic absorbers detected in the 7 QSO spectra. Square brackets denote averages of the (similar) velocities measured from more than one line.}
\end{table}

 We identify a total of 12 candidate intrinsic absorbers in the RL-QSO spectra (see Tables 3 to 8 and the compilation in Table 9), the majority (9/12) of these blueshifted relative to the AGN, with a mean $\Delta(v)$ of -249 km$\rm ^{-1}$ and a wide spread $\sigma(\Delta(v))=712$ km $\rm s^{-1}$. This is general agreement with findings for HzRGs (van Ojik et al. 1997), which can similarly
have very  (by $>600$ km $\rm s^{-1}$) blueshifted absorbers (Humphrey et al. 2008).

  Of these 12 absorbers, 7 are seen only as  Lyman-$\alpha$ absorption, while for 4 there are CIV and/or NV and/or SiIV absorption lines as well as Lyman-$\alpha$ (with the same $\Delta(v)$), indicating a particularly high ionization of the absorbing gas (as previously seen for some absorbers in HzRGs and RL-QSO spectra; e.g. Binette et al. 2000, Humphrey et al. 2013), and one is SiII1260.
 
 One absorption line, in Q1658+575, may be visible  on the  red side of the Lyman-$\alpha$ emission from the nebula as well as in the QSO spectrum (Fig 4) and could therefore be an extended absorber. The strong Lyman-$\alpha$ absorption line bisecting the Q0805+046 Lyman-$\alpha$ may also affect the nebular spectrum (Fig 8) but would appear to have  a lower covering factor than for the  QSO. Such candidate extended absorbers could be confirmed using integral-field spectroscopy to map the spatial extent of absorption against the bright background of nebular Lyman-$\alpha$ emission.

\section{Summary of Conclusions}
(i) We performed long-slit spectroscopy of 6 radio-loud QSOs at redshifts $2<z<3$, with OSIRIS on the Gran Telescopio Canarias, targeting the Lyman-$\alpha$ line. The QSOs were selected from Heckman et al. (1991a) as having  strong Lyman-$\alpha$ emission from surrounding nebulae (HzLAN) extending to radii of tens of kpc.  For 4/6 we detect the giant nebula as an extension of the QSO's Lyman-$\alpha$ line for 1--9 arcsec (8--72 kpc) outside the AGN spectrum, on one or both sides of the slit (one non-detection might be due to very poor observing conditions, and the other, non-optimal slit PA plus low sensitivity at the blue extreme of the $\lambda$ range).
\medskip

(ii) With the same instrument we obtained the long-slit spectrum of a much higher redshift radio-loud QSO, SDSS J2228+0110 at $z\simeq 5.9$. We find evidence its Lyman-$\alpha$ emission extends at least 1--2 arcsec (6--12 kpc) beyond the AGN continuum spectrum, indicating a giant  Lyman-$\alpha$ emission nebula, which is a new discovery for this galaxy. On the basis of this and 2 other known examples, QSOs and starburst galaxies at $z\sim 6$ can have giant Lyman-$\alpha$ emitting nebulae of equal intrinsic (i.e. correcting for redshifting) surface brightness to their counterparts at $2<z<3$ .
 \medskip

(iii) The spectra of the 5 nebulae show Lyman-$\alpha$ emission lines with surface brightness  0.4--$3\times 10^{-16}$  ergs $\rm cm^{-2}s^{-1}arcsec^{-2}$, and FWHM 400--1100 (mean 863) km $\rm s^{-1}$ (tending to become narrower at greater distances from the central galaxy). For one nebula (Q2222+046) we detect a second, much fainter (by factor 18) and narrower emission line at $5449\rm \AA$ and identify this as HeII1640.
\medskip
 
 (iv) In the QSO spectra we also identify a total of 12 probable intrinsic absorbers: 7 only as Lyman-$\alpha$ absorption, 4 also in NV, CIV and/or SiIV, and one example of SiII1260. It is not always possible to determine whether intrinsic Lyman-$\alpha$  absorption extends across the nebula, as the nebula's emission line is always narrower than the QSO's, but  there is some indication of this for a line in Q1658+575, and possibly to some extent in Q0805+046.
\medskip

(v) We examine the spatially-resolved kinematics of the extended nebulae by fitting to their Lyman-$\alpha$ line at a series of positions along the slit. The line-of-sight $\Delta(v)$ profiles were generally almost flat across the nebulae, in contrast to many radio galaxy nebulae which show large gradients. However, 3 of the 5 nebulae were systematically redshifted relative to the Lyman-$\alpha$ line of the QSO, by $\Delta(v)=250$--460 km $\rm s^{-1}$ (giving V-shaped velocity profiles which flatten at larger radii). We consider the most likely interpretation to be that  the motion of this extended gas is dominated by infall, which would be consistent with the results of Humphrey et al. (2007, 2013) and Villar-Mart\'{i}n et al. (2007), who reached similar conclusions for type 2 radio-loud AGNs at $z>2$.  For one of these nebulae, Q0805+046, the brighter side appears to be slightly ($\Delta(v)\simeq 100$ km $\rm s^{-1}$) more redshifted and we also found a steep velocity gradient ($\sim 400$ km $\rm s^{-1}$), across the central 2 arcsec. The other two nebulae, including the $z=5.9$ example, showed no significant offset from the AGN in mean velocity.
\medskip

(vi) Giant Lyman-$\alpha$ nebulae associated with radio galaxies and with radio-loud QSOs at $2<z<3$ appear to be very similar in most properties including luminosity, size and line FWHM. However our analysis suggests a systematic difference in the velocity maps. This could simply reflect the different orientation of these two classes of source to the observer's line of sight, resulting in a component of infall of the nebula (rather than systematic rotation) being brought into greater prominence.
\medskip

\section*{Acknowledgements}
This work is based on observations made during Mexican time with the Gran Telescopio Canarias at the Observatorio del Roque de los Muchachos, La Palma, Spain. NR acknowledges the support of `Funda\c{c}\~ao para a Ci\^encia e a Tecnologia' Investigador grant SFRH/BI/52155/2013. AH acknowledges a Marie Curie Fellowship co-funded by the 7$^{th}$  Research Framework Programme and the Portuguese Funda\c{c}\~ao para a Ci\^encia e a Tecnologia. AH acknowledges support by the Funda\c{c}\~ao para a Ci{\^e}ncia e a Tecnologia (FCT) under project FCOMP-01-0124- FEDER-029170 (Reference FCT PTDC/FIS-AST/3214/2012), funded by FCT- MEC (PIDDAC) and FEDER (COMPETE). LB acknowledges CONACyT grant 128556. This research has made use of the NASA/IPAC Extragalactic Database (NED) which is operated by the Jet Propulsion Laboratory, California Institute of Technology, under contract with the National Aeronautics and Space Administration. We also thank Montse Villar-Mart\'in, Itziar Aretxaga and Polychronis Papaderos for useful discussions.

\section*{References} 

\vskip0.15cm \noindent  Adams J.J., Hill G. J., MacQueen P.J., 2009, ApJ, 694, 314. 	

\vskip0.15cm \noindent  Barthel P.D., Miley, G.K., Schilizzi R.T., Lonsdale C.J., 1988, 
A\&AS 73, 515.

\vskip 0.15cm  \noindent Bicknell G., Sutherland R., van Breugel W., Dopita M., Dey A., Miley G., 
2000, ApJ 540, 678.

\vskip 0.15cm  \noindent Binette L., Kurk J.D., Villar-Mart\'in M., R\"ottgering, H.J.A., 2000, A\&A 356, 23.

\vskip 0.15cm  \noindent Cantalupo S., Arrigoni-Battaia F., Prochaska J.X., Hennawi J.F., Madau P.,  2014, Nature 506, 63.

\vskip0.15cm \noindent Christensen L., Jahnke K., Wisotzki L., and S\'anchez S.F., 2006, A\&A 459, 717.

\vskip0.15cm \noindent Colbert J.W., Scarlata C., Teplitz H., Francis P., Palunas P., Williger G.M., Woodgate B., 2011, ApJ 728, 59.

\vskip0.15cm \noindent Dekel A., Birnboim Y., Engel G., Freundlich J., Goerdt T., Mumcuoglu M., Neistein E., Pichon C., Teyssier R., Zinger E.,  2009, Nature 457, 451.

\vskip0.15cm \noindent Dijkstra M., Haiman Z.,  Spaans M., 2006, ApJ 649, 37.

\vskip0.15cm \noindent Goerdt T., Dekel A., Sternberg A., Ceverino C.,Teyssier R., Primack J.R.,
2010, MNRAS, 407, 613.

\vskip0.15cm \noindent Goto T., Utsumi Y., Furusawa H., Miyazaki S., Komiyama Y., 2009, MNRAS 400, 843.
 
 \vskip0.15cm \noindent Goto T., Utsumi Y., Walsh J.R., Hattori T., Miyazaki S.; Yamauchi C., 2012, MNRAS 421, L77.
  	
\vskip0.15cm \noindent Hayes, M.,  Scarlata C., Siana B., 2011, Nature 476, 304.

\vskip0.15cm \noindent Heckman T.M., Lehnert M.D., van Breugel W., Miley G.K., 1991a,
ApJ 370, 78. (H91)

\vskip 0.15cm  \noindent Heckman T.M., Lehnert M.D., Miley G.K., van Breugel W., 1991b,
ApJ 381, 373. 

\vskip 0.15cm  \noindent Humphrey A., Villar-Mart\''n M., Fosbury R., Vernet J., di Serego Alighieri S.,  2006, MNRAS 369, 1103.

\vskip 0.15cm  \noindent Humphrey A., Villar-Mart\'in M., Fosbury R., Binette L., Vernet J., De Breuck C., di Serego Alighieri, S., 2007, MNRAS 375, 705.

\vskip 0.15cm  \noindent Humphrey A., Villar-Mart\'in, M., S\'anchez S.F., di Serego Alighieri S., De Breuck C., Binette L., Tadhunter C., Vernet J., Fosbury R., Stasielak J,  2008, MNRAS 390, 1505.

\vskip 0.15cm  \noindent Humphrey A., Iwamuro F., Villar-Mart\'in M. Binette L., Sung E.C., 2009, MNRAS 399, L34.

\vskip 0.15cm  \noindent Humphrey A., Binette L., Villar-Mart\'in M., Aretxaga I., Papaderos P.,
2013, MNRAS 428, 563.

\vskip 0.15cm  \noindent Humphrey A., Vernet J., Villar-Mart\'in M., di Serego Alighieri S., Fosbury R.A.E., Cimatti A., 2013b, ApJ 768, L3,

\vskip 0.15cm  \noindent  Lehnert M.D, Becker R.H., 1998, A\&A 332, 514.

\vskip 0.15cm  \noindent  Lehnert M.D., van Breugel W.J.M., Heckman T.M., Miley G.K., 1999,
 ApJSS 124, 11.

\vskip 0.15cm  \noindent  Martin D.C., Chang D., Matuszewski M., Morrissey P., Rahman S., Moore A., Steidel C.C., 2014, ApJ 786, 106.

\vskip 0.15cm  \noindent Nesvadba N.P.H., Lehnert M.D., De Breuck C., Gilbert A.M., van Breugel W.,
2008, A\&A 491, 407.	

\vskip 0.15cm  \noindent van Ojik R., R\"ottgering H.J.A., Miley G.K., Hunstead R.W., 1997, 
A\&A 317, 358.

\vskip 0.15cm  \noindent  Ouchi M., et al., 2009, ApJ 696, 1164.

\vskip 0.15cm  \noindent  Ouchi M., et al., 2013, ApJ 778, 102.

\vskip 0.15cm  \noindent  Roche N.D., Franzetti P., Garilli B., Zamorani G., Cimatti A., Rossetti E., 2012, MNRAS 420, 1764.

\vskip 0.15cm  \noindent  S\'anchez S.F., Humphrey A., 2009, A\&A 495, 471.

\vskip 0.15cm  \noindent S\'anchez Almeida J., Elmegreen B. G., Mun\~oz-Tun\~on C., Elmegreen D., M., 2014, A\&A preprint (arXiv1405.3178).

\vskip 0.15cm  \noindent Schaerer D., Hayes M., Verhamme A., Teyssier, R.,  2011, A\&A, 531, 12.

\vskip 0.15cm  \noindent Smith D.J.B., Jarvis M. J., Simpson C., Mart\'inez-Sansigre A., 2009, MNRAS 393, 309.

\vskip 0.15cm  \noindent Stockton A., McGrath E., Canalizo G., Iye M., Maihara T., 2008, ApJ 672, 146.

\vskip 0.15cm  \noindent Verhamme A., Schaerer D., Atek H., Tapken C.,  2008, A\&A, 491, 89.

\vskip0.15cm  \noindent Villar-Mart\'in M., Vernet J., di Serego Alighieri S., Fosbury R., 
Humphrey A., Pentericci L., 2003, MNRAS, 346, 273.

\vskip0.15cm  \noindent Villar-Mart\'in M., S\'anchez S. F., De Breuck C., Peletier R., Vernet J., Rettura A., Seymour N., Humphrey A., Stern D., di Serego Alighieri S., Fosbury R., 2006, 
MNRAS 366, L1.

\vskip0.15cm  \noindent  Villar-Mart\'in M., S\'anchez S.F., Humphrey A., Dijkstra M., di Serego Alighieri S., De Breuck C., Gonz\'alez Delgado R., 2007, MNRAS 378, 416.

\vskip 0.15cm  \noindent Villar-Mart\'in M., Humphrey A., Delgado R.G., Colina L., Arribas S., 2011, 
MNRAS 418, 2032.

\vskip 0.15cm  \noindent Weidinger M., M\.oller P., Fynbo J.P.U., Thomsen B., 2005, A\&A 436. 825.

\vskip 0.15cm  \noindent Willott C.J., Chet S., Bergeron J., Hutchings J.B., 2011,  AJ 142, 186.

\vskip 0.15cm  \noindent Yajima H., Li Y., Zhu Q., 2013, ApJ 773, 151.

\vskip 0.15cm  \noindent Yang Y., Walter F., Decarli R., Bertoldi F., Weiss A., Dey A., Prescott M.K.M., Badescu T., 2014, ApJ 784, 171.

\vskip0.15cm  \noindent  Zeimann G.R., White R.L., Becker R.H., Hodge J.A., Stanford S.A., Richards G.T., 2011, ApJ 736, 57.

\end{document}